\documentclass[a4paper]{article}

\pdfoutput=1
\usepackage[utf8]{inputenc}

\usepackage[english]{babel}
\usepackage[T1]{fontenc}
\usepackage{csquotes}

\usepackage[a4paper,top=3cm,bottom=2cm,left=3cm,right=3cm,marginparwidth=1.75cm]{geometry}

\usepackage{amsmath}
\usepackage{graphicx}
\usepackage{hyperref}
\hypersetup{colorlinks=true, allcolors=blue}
\usepackage{authblk}
\usepackage{multirow}
\usepackage{tabularx}
\usepackage{makecell}
\usepackage{hyperref}
\usepackage[backend=biber, sorting=none]{biblatex}
\usepackage{nomencl}
\usepackage{booktabs}
\usepackage{tabularx}
\usepackage{textcomp} 
\usepackage{amsfonts}
\usepackage{outlines} 
\usepackage{subfig}
\usepackage[version=4]{mhchem}
\usepackage{siunitx}
\usepackage{bm}
\usepackage[normalem]{ulem} 
\date{}

\DeclareSIUnit\angstrom{\text {Å}}
\usepackage{parskip}
\usepackage[final]{changes}

\title{Generative models for crystalline materials}

\author[1,2,${\dagger}$]{Houssam Metni}
\author[2,${\dagger}$]{Laura Ruple}
\author[3,4]{Lauren N. Walters}
\author[1,2]{Luca Torresi}
\author[1,2]{Jonas Teufel}
\author[2]{Henrik Schopmans}
\author[1,2]{Jona Östreicher}
\author[1,2]{Yumeng Zhang}
\author[1,2]{Marlen Neubert}
\author[1,2]{Yuri Koide}
\author[2]{Kevin Steiner}
\author[2]{Paul Link}
\author[2]{Lukas Bär}
\author[2,$+$]{Mariana Petrova}
\author[3,4]{Gerbrand Ceder}
\author[1,2,*]{Pascal Friederich}

\affil[1]{Institute of Nanotechnology, Karlsruhe Institute of Technology, Kaiserstr. 12, 76131 Karlsruhe, Germany}
\affil[2]{Institute of Anthropomatics and Robotics, Karlsruhe Institute of Technology, Kaiserstr. 12, 76131 Karlsruhe, Germany}
\affil[3]{Department of Materials Science and Engineering, University of California, Berkeley, California 94720, United States}
\affil[4]{Materials Sciences Division, Lawrence Berkeley National Laboratory, California 94720, United States}

\affil[$+$]{Current address: INIT GmbH, Käppelestraße 4-10, 76131 Karlsruhe, Germany.}
\affil[$\dagger$]{These authors contributed equally.}

\affil[*]{Corresponding author: pascal.friederich@kit.edu}

\begin{document}

\maketitle
\begin{abstract}
Understanding structure-property relationships in materials is fundamental in condensed matter physics and materials science. Over the past few years, machine learning (ML) has emerged as a powerful tool for advancing this understanding and accelerating materials discovery.
Early ML approaches primarily focused on constructing and screening large material spaces to identify promising candidates for various applications. 
More recently, research efforts have increasingly shifted toward generating crystal structures using end-to-end generative models.
This review analyzes the current state of generative modeling for crystal structure prediction and \textit{de novo} generation. It examines crystal representations, outlines the generative models used to design crystal structures, and evaluates their respective strengths and limitations. 
Furthermore, the review highlights experimental considerations for evaluating generated structures and provides recommendations for suitable existing software tools. Emerging topics, such as modeling disorder and defects, integration in advanced characterization, incorporating synthetic feasibility constraints, and model explainability are explored. Ultimately, this work aims to inform both experimental scientists looking to adapt suitable ML models to their specific circumstances and ML specialists seeking to understand the unique challenges related to inverse materials design and discovery.
\end{abstract}

\section{Introduction}

The discovery of new materials has been one of the most important drivers of technological development. Over time, efforts to address the materials discovery challenge have progressed through experiment-driven advances, the development of theoretical frameworks, and large-scale simulations. Today, these approaches are complemented by a fourth paradigm~\cite{agrawal2016perspective}: data-driven science, where machine learning (ML) approaches, powered by extensive databases of experimental and simulated data~\cite{hellenbrandt2004inorganic, Jain2013, draxl2018nomad, Scheffler2022, barroso2024open, schmidt2024improving, dagdelen2024structured} as well as modern computing infrastructure, open unprecedented opportunities for materials design~\cite{sanchez2018inverse, noh_inverse_2019, schmidt2019recent, Speckhard2025}.

Within the fourth paradigm, a variety of computational methods have been developed to explore and design crystal structures with application-specific properties. Conventional high-throughput virtual screening approaches were enhanced with active learning approaches to accelerate the generation of large informative sets of hypothetical materials using combinations of density functional theory (DFT), predictive ML methods, and uncertainty estimation~\cite{gomez2016design}. Complementing this, machine-learned interatomic potentials (MLIPs) accelerate optimization and sampling workflows and thus enable fast property estimation and efficient down-selection from massive candidate spaces~\cite{GNOME_Merchant2023}. Evolutionary and stochastic search algorithms~\cite{GNOME_Merchant2023}, as well as rule-based structure generators, provide additional strategies to systematically explore structural possibilities, often guided by physical constraints or heuristic rules. 
Furthermore, the development of automated high-throughput labs~\cite{potyrailo2011combinatorial, green2017fulfilling, ludwig2019discovery} as well as increasingly autonomous materials acceleration platforms and self-driving labs~\cite{Canty2025sdl, Dai2024autonomous, szymanski2023autonomous, abolhasani2023rise, tom2024self, maffettone2023missing} are starting to close the gap between virtual and experimental materials design and discovery.
Together, these developments illustrate a clear methodological progression: from physics-based simulations to ML-augmented searches, towards autonomous virtual and experimental discovery, generating increasing volumes of ML-ready data, which in turn accelerates the design and discovery process.

One of the key components in this evolution in methodology is end-to-end generative models, which aim to directly propose material structures without relying on exhaustive searches or handcrafted rules~\cite{sanchez2018inverse}. By learning structural patterns and symmetries from data, generative approaches invert the traditional workflow: instead of generating candidates and evaluating them sequentially, they seek to directly suggest new structures that satisfy desired constraints from the outset. Over the past ten years, new developments in generative models from the core machine learning community were quickly adopted for inverse design of molecules and crystal structures. Compared to generative models for molecules~\cite{bilodeau2022generative}, crystal generation presents additional challenges: crystals are periodic, often involve complex unit cells, and must respect strict symmetry constraints, making representation and model design substantially more intricate. For this reason, transfer of new approaches to crystal generation was usually delayed by 1-2 years. For instance, the first successful attempts to use variational autoencoders (VAEs) for images were published in 2013~\cite{kingma2013auto}, then adapted to generate molecular structures in 2016~\cite{gomez2018automatic}, and to materials in 2019~\cite{noh_inverse_2019}. Generative adversarial networks (GANs), introduced in 2014~\cite{goodfellowGenerativeAdversarialNetworks2014}, were first used for molecules in 2018~\cite{de2018molgan}, for materials compositions in 2018~\cite{nouiraCrystalGANLearningDiscover2019}, and for crystal structures in 2020~\cite{kimGenerativeAdversarialNetworks2020}. Normalizing flows, introduced in 2014~\cite{nice}, were applied to molecules in 2019~\cite{madhawa2019graphnvp} and to crystal structures in 2022 and 2023~\cite{wirnsbergerNormalizingFlowsAtomic2022, kohlerRigidBodyFlows2023}. The same is true for diffusion models (introduced in 2020~\cite{ho2020denoising}, applied to molecules in 2021 and 2022~\cite{shi2021learning, xu2022geodiff, hoogeboom2022equivariant}, and to crystal structures in 2022~\cite{xie2021crystal}), and transformers (introduced in 2017~\cite{vaswani2017attention}, first applied to molecules in 2019-2021~\cite{schwaller2019molecular, bagal2021molgpt}, and to crystal structures in 2023 and 2024~\cite{flam-shepherd2023LanguageModelsCan, antunes2024CrystalStructureGeneration}).

Recent reviews have broadly discussed and offered perspectives on ML-assisted inverse design in materials science~\cite{Han2025, handoko2025artificialintelligencegenerativemodels, ParkWalsh2024, wang2025crystallinematerialdiscoveryera, Cheng2026}, some of which have focused specifically on crystal structure prediction tasks~\cite{Li2024, Chen2025, Wang2025}.
\citeauthor{debreuck2025generativeaicrystalstructures} provided a comprehensive and technical overview of the methods and models used in generative approaches for crystal structures, with discussions on methodologies and future directions, but with a primary focus on the ML models themselves~\cite{debreuck2025generativeaicrystalstructures}.
\citeauthor{li2025materialsgenerationeraartificial} similarly emphasized the models, while also discussing datasets and evaluation metrics, and offering a brief perspective on experimental implementation, highlighting challenges such as synthesizability~\cite{li2025materialsgenerationeraartificial}.
More recently, \citeauthor{Park2025synthesisgap} provided a perspective on the need to integrate synthesizability considerations into computational materials design~\cite{Park2025synthesisgap}. This perspective article emphasized that effective inverse design must account not only for predicted stability but also for practical pathways to experimental realization, a topic of increasing importance that will also be reviewed here.

In this work, we extend these efforts in multiple aspects by focusing on generative models for crystalline materials with (\textit{i}) an emphasis on recent innovations in ML approaches in the last years, \textit{i.e.}, models based on diffusion, flows, and transformers; (\textit{ii}) a specific focus on introducing conceptual differences of representations, datasets, and evaluation metrics, also making these distinctions accessible to non-experts and new researchers in this field; and (\textit{iii}) to enhance practical relevance beyond previous review papers, a discussion of the practical adaptation of generative models for experimental discovery, as well as the current limitations and emerging topics, \textit{e.g.}, defects and disorder, characterization workflow integration, synthetic feasibility, and model interpretability.\\
Combining these three aspects, this review should serve as a useful guide for an audience of experimental and computational materials scientists looking to find an entry point to existing generative models and how to use them in their specific applications and related materials design challenges. At the same time, this review should also inspire the materials informatics and computer science communities, looking for an experimental perspective on currently open challenges and considerations to take into account in the next generation of generative models for materials.
\section{Representations}

\begin{figure}
    \centering
    \includegraphics[width=0.95\linewidth]{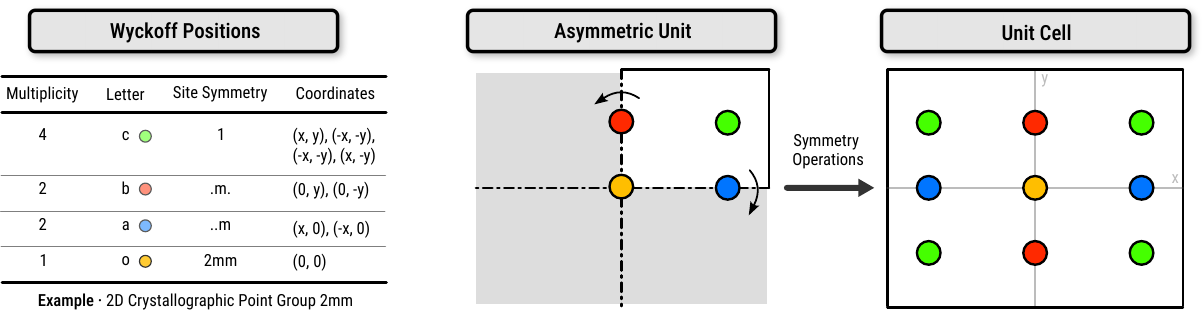}
    \caption{Overview of the Wyckoff positions, asymmetric units, and symmetry operations in a unit cell.}
    \label{fig:figure_5}
\end{figure}

\begin{figure}[b!]
    \centering
    \includegraphics[width=0.95\linewidth]{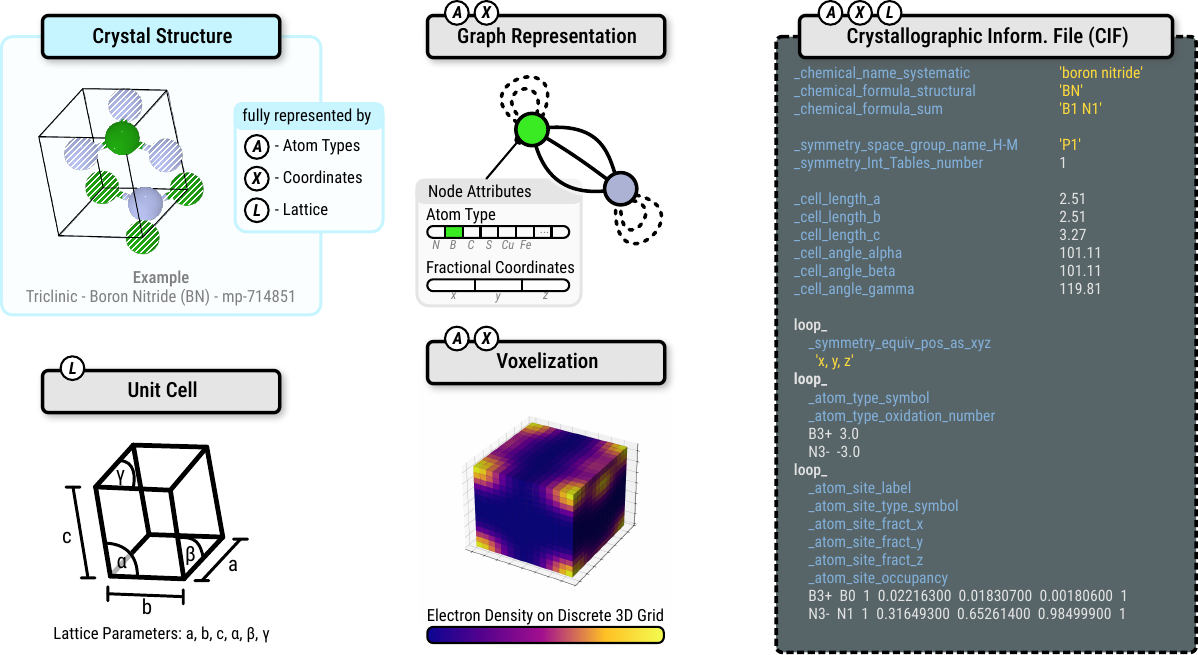}
    \caption{Overview of the unit cell and machine-readable representations of crystal structures.}
    \label{fig:figure_1}
\end{figure}
%


Crystals are solids composed of atoms, ions, or molecules arranged in a periodic, three-dimensional pattern. This periodicity is described using a unit cell, the smallest repeating unit that fully captures the symmetry and structure of the crystal. Unit cells are defined by lattice parameters—lengths $a$, $b$, $c$ and angles $\alpha$, $\beta$, $\gamma$—and contain a basis of atoms. Translating the unit cell through space along lattice vectors generates the infinite crystal lattice.

Symmetry plays a fundamental role in crystallography (\textbf{Figure \ref{fig:figure_5}}). Crystals are categorized into space groups, which define all symmetry operations (including rotations, reflections, inversions, and fractional translations) that leave the crystal structure invariant~\cite{InternationalCryst}. Each space group also defines an asymmetric unit, the smallest portion of the structure that can be used to reconstruct the entire unit cell through the application of the group’s symmetry operations. Atomic positions in the unit cell are described using Wyckoff positions, which specify the symmetry-allowed coordinates of atoms. Each Wyckoff position is associated with a site multiplicity (the number of equivalent positions generated by symmetry) and a site symmetry (the local point group symmetry of that position). This formalism allows crystal structures to be described compactly, while strictly adhering to symmetry constraints.
Generative models for crystal structures employ a variety of machine-readable representations, such as the Crystallographic Information File (CIF)~\cite{hall1991crystallographic}, graph-, and voxel-based formats (\textbf{Figure \ref{fig:figure_1}}). Additionally, some models also enforce space group symmetry constraints during training or generation, ensuring that generated structures adhere to crystallographic symmetry rules.

CIF is the standard format for storing and exchanging crystal structure data. CIF files contain comprehensive details about unit cell parameters, symmetry information (space group and symmetry operations), atomic positions (fractional coordinates, occupancy), and often Wyckoff positions, either explicitly or implicitly. Additional metadata from experimental or computational sources is typically included as well. CIF files are widely used in crystallographic databases and serve as a standard input/output format for many tools in generative modeling.

A more flexible and machine-learning-friendly representation of crystal structures is through graphs~\cite{CGCNN_Xie_2018}. In graph-based models, nodes correspond to atoms and are represented with necessary features such as element type (\textit{e.g.}, atomic number) and additional information such as electronegativity and valence. Edges represent interatomic interactions or spatial proximity, typically determined using distance-based cutoffs or k-nearest neighbors~\cite{ruff2024connectivity}. Periodic boundary conditions are incorporated by including edges that span across unit cell boundaries, effectively capturing the infinite periodic nature of crystals. These representations preserve the spatial and chemical relationships within the structure and are well-suited for use with graph neural networks (GNNs). Furthermore, symmetry can be encoded directly into the graph model: translational symmetry is handled through periodic boundary conditions, while rotational and reflectional symmetries can be incorporated via equivariant neural networks (\textit{e.g.}, E(3)-equivariant GNNs or SE(3)-Transformers), which ensure that the model’s outputs transform consistently under these operations~\cite{satorras2021n, SE3Transformer}.

Crystal structures can also be represented using voxelization, a technique in which the 3D space of the unit cell is discretized into a regular grid of volumetric pixels (voxels). Each voxel encodes physical information, such as atomic occupancy, electron density, or atomic species, within that spatial region. The resulting 3D tensor representation is analogous to a 2D image and can be processed by 3D convolutional neural networks (CNNs) or volumetric generative models~\cite{longConstrainedCrystalsDeep2021}. Voxel-based representations, when processed with fully-convolutional networks, are translation-equivariant. A shift in the input produces a corresponding shift in the feature maps. True translation invariance is only obtained if the architecture removes spatial information (\textit{e.g.}, via pooling). Rotational invariance can be approximated with data augmentation. However, due to the sparseness of the unit cell, voxel methods are computationally expensive and resolution-limited, making them less common than graph-based approaches for generative modeling tasks.

It is worth noting that alternative crystal structure representation methods have recently been proposed, although they have not yet been applied in the context of discriminative or generative ML models.
For instance,~\citeauthor{Mrdjenovich2024} introduced an algorithm representing each crystal as a unique set of integers derived from its lattice and atomic basis~\cite{Mrdjenovich2024}. The lattice is encoded through an obtuse superbasis whose geometric features are discretized, while atomic positions are binned and ordered to remove symmetry and permutation redundancies. Together, these yield a canonical ``crystal normal form”, enabling efficient comparison and exploration of relationships between different crystal structures. Exploring new representations in generative models might increase performance without requiring more data or innovations in the model architectures.
\section{Databases}

The effectiveness of ML methods strongly depends on the quality of the underlying data, as high-quality data is crucial for ensuring model reliability and generalization. The development of comprehensive and accurate crystal structure databases is thus of paramount importance in computational materials science, motivating the community efforts to invest in curated databases.
On the one hand, the experimental materials science community has been gathering measurements and characterization results into open-access databases. On the other hand, the scarcity of experimental data has motivated the generation of large-scale simulated datasets using DFT, which complement and expand existing resources. While these databases are often categorized as experimental or DFT-calculated, many are hybrid in practice. Additional efforts have also led to the creation of application-specific databases tailored to particular materials.
It is important to note that existing crystal structure databases are inherently imbalanced. Certain chemical elements, space groups, and structure types are strongly overrepresented due to historical research focus, experimental accessibility, and computational screening strategies. These representation biases can influence generative models in the absence of additional constraints. However, the quantitative impact of these biases remains difficult to assess given the lack of alternative, unbiased reference datasets.
In this section, we present existing publicly available crystal structure databases, also summarized in \textbf{Table~\ref{Table_1}}.

\begin{table}[htbp]
\centering
\renewcommand{\arraystretch}{2}
\resizebox{\textwidth}{!}
{%
\begin{tabular}{|l|l|l|l|l|}
\hline
\textbf{Name \& reference} & \textbf{Experimental validation} & \textbf{Database size} & \textbf{Material types}\\
\hline
Materials Project (Full)~\cite{Jain2013} & Mixed (some verified) & \textgreater{}700,000 total,~200,000 crystals & Molecules and inorganic crystals \\
\hline
Materials Project (MP-20)~\cite{xie2021crystal} & Yes & $\sim$45,000 & Inorganic crystals\\
\hline
Materials Project (MPTS-52)~\cite{levy2025SymmCDSymmetryPreservingCrystal} & Yes & $\sim$40,000 & Inorganic crystals \\
\hline
ICSD~\cite{zagorac2019recent} & Mixed (mostly verified) & \textgreater{}320,000  & Inorganic crystals \\
\hline
CSD~\cite{Groom2016} & Yes & 1.25 million  & Organic and metal-organic crystals\\
\hline
COD~\cite{gravzulis2009crystallography} & Mixed (less filters) & \textgreater{}520,000 & Organic, inorganic, metal-organic crystals \\
\hline
AFLOW~\cite{curtarolo2012aflow} & Computational & \textgreater{}3.9 million  & Inorganic crystals\\
\hline
Alexandria~\cite{ghahremanpour2018alexandria}& Computational & \textgreater{}5 million  & Molecules, inorganic and low-dimensional materials \\
\hline
JARVIS-DFT~\cite{choudhary2020joint} & Computational & $\sim$41,000  & 0D to 3D materials \\
\hline
OQMD~\cite{kirklin2015open} & Mixed (some verified) & \textgreater{}1.3 million  & Inorganic crystals \\
\hline
C2DB~\cite{haastrup2018computational} & Mixed (some verified) & \textgreater{}4,000 & 2D materials \\
\hline
2DMatPedia~\cite{zhou20192dmatpedia} & Computational & \textgreater{}6,000  & 2D materials \\
\hline
Carbon-24~\cite{pickard2020airss} & Computational & $\sim$10,000  & Carbon structures \\
\hline
Perov-5~\cite{castelli2012new,castelli2012computational} & Computational & $\sim$19,000  & Perovskite structures \\
\hline
OC20~\cite{chanussot2021open} & Computational & \textgreater{}1.2 million  & Catalysts \\
\hline
NEMAD~\cite{zhang2024gptarticleextractor, itani2024northeast} & Yes & $\sim$67,000  & Magnetic materials \\
\hline
SuperCon~\cite{materials-a} & Yes & $\sim$33,000  & Superconductors \\
\hline
3DSC~\cite{sommer20233dsc} & Yes & $\sim$9,150 & Superconductors \\
\hline
NOMAD~\cite{draxl2018nomad, Sbailo2022, Scheidgen2023} & Mixed (some verified)& \textgreater{}19 million  & A wide range of materials\\
\hline
\end{tabular}
}
\caption{Summary of major publicly available materials databases from the literature, highlighting key characteristics including database size and material types covered. Experimental validation is indicated where applicable — while some databases are purely computational, others combine theoretical and experimentally verified structures to varying degrees.}
\label{Table_1}
\end{table}

\subsection{Experimental crystal structure databases}
The Inorganic Crystal Structure Database (ICSD), managed by the Leibniz Institute for Information Infrastructure (FIZ) Karlsruhe, is one of the largest databases for over 320{,}000 completely identified inorganic crystal structures, all of which have undergone quality checks to be verified~\cite{zagorac2019recent}.
The Cambridge Structural Database (CSD), managed by the Cambridge Crystallographic Data Centre (CCDC), consists of 1.25 million entries of small-molecule organic and metal-organic crystal structures, mainly data from X-ray and neutron diffraction analyses, spanning various experimental applications~\cite{Groom2016}. 
The Crystallographic Open Database (COD) is a collection of 520{,}000 small to medium-sized unit cell crystal structures of organic, inorganic, metal-organic compounds, and minerals~\cite{gravzulis2009crystallography}.
Due to the open-access contribution model of this database, there are fewer filters to verify the validity and stability of the added structures.
The Predicted Crystallographic Open Database (PCOD) is a COD subset of crystal
structures predicted by focusing on predicted structures generated through computational methods such as GRINSP~\cite{LeBail_cg5019}, while COD encompasses both experimentally determined and computationally predicted structures. 

\subsection{Databases with DFT-calculated properties}

Beyond the structure and potential accompanying experimental analyses, some databases also calculate further properties using DFT.
The Materials Project (MP) is an open database containing structural, electronic, and energetic properties of known and predicted materials, including over 500{,}000 molecules and 200{,}000 inorganic materials~\cite{Jain2013}. 
A subset of the MP consisting of inorganic materials with less than 20 atoms in the unit cell, labeled as MP-20, contains about 45,000 metastable crystal structures spanning 89 different element types~\cite{xie2021crystal}.  
The Materials Project time split data (MPTS-52) is another subset of the MP comprising experimentally verified compounds with less than 52 atoms per unit cell~\cite{levy2025SymmCDSymmetryPreservingCrystal}.

The AFLOW (Automatic FLOW for Materials Discovery) repository encompasses data on approximately 3.9 million material compounds, along with a vast array of computed materials electronic, thermal, mechanical and thermodynamic properties~\cite{curtarolo2012aflow}.
Similarly, the Alexandria library is a quantum-chemical database of inorganic crystals, molecules, and low-dimensional materials whose geometries were optimized using over 5 million DFT calculations~\cite{ghahremanpour2018alexandria}. 
Other notable databases and benchmark tasks include the JARVIS-DFT~\cite{choudhary2020joint}, OQMD~\cite{kirklin2015open}, Meta's OMAT24~\cite{barroso2024open}, NOMAD~\cite{draxl2018nomad, Sbailo2022, Scheidgen2023}, as well as the C2DB~\cite{haastrup2018computational}, 2DMatPedia~\cite{zhou20192dmatpedia}, both with a focus on two-dimensional (2D) materials.

\subsection{Application-specific databases}

The Carbon-24 dataset focuses on 10{,}153 carbon structures selected from an original set of 101{,}529 structures~\cite{pickard2020airss}. These share the same elemental composition but vary in structure, spanning 6 to 24 atoms per unit cell. The Perov-5 dataset comprises properties of about 19{,}000 perovskite materials, curated from a dataset originally intended for water-splitting~\cite{castelli2012new,castelli2012computational}. 
In both of the previous datasets, the structures considered are not necessarily thermodynamically stable, so they might require further pre-selection protocols to filter the experimentally synthesizable and stable structures. 

For catalysts, the OC20 dataset consists of 1{,}281{,}040 DFT relaxations (264{,}890{,}000 single-point evaluations) across a wide swath of materials, surfaces, and adsorbates (nitrogen, carbon, and oxygen chemistries)~\cite{chanussot2021open}.
The Northeast Materials Database (NEMAD) is a database that consists of 67{,}573 magnetic materials collected via LLM-based automated scientific data extraction from experimental literature~\cite{zhang2024gptarticleextractor,itani2024northeast}. 
Supercon focuses on high-$T_c$ oxide superconductors, metal-alloy superconductors, and organic superconductors, with data extracted from academic papers and other sources~\cite{materials-a}, complemented by 3DSC, a dataset of superconductors based on Supercon but including matched crystal structures from the ICSD and Materials Project~\cite{sommer20233dsc}.

\section{Crystal structure related tasks, generation approaches, and machine learning models}

Crystal structure–related tasks can be broadly categorized into two main types: property prediction and structure generation. Property prediction is a ``forward task" which involves estimating characteristics such as formation energy, band gap, or mechanical stability for a given crystal structure, using methods ranging from classical first-principles approaches like DFT to modern predictive ML models such as GNNs and large language models (LLMs).
When combined with large-scale computational screening, predictive approaches enable the rapid exploration of vast materials libraries.
In contrast, structure generation is an ``inverse task", focusing on designing or generating new crystal structures that satisfy specific criteria. This includes tasks such as predicting stable crystal structures from chemical composition and generating entirely new crystal structures from scratch.
Crystal structure prediction (CSP) has long been of paramount importance for understanding materials' properties and designing them. 
It consists of determining the most stable atomic arrangement for a given chemical composition.
\textit{De novo} generation aims to go a step further by creating entirely new crystal structures, potentially with novel compositions, from scratch. 
While CSP is constrained to exploring structural possibilities within a fixed chemical formula, typically relying on search algorithms and energy evaluations to identify low-energy configurations, \textit{de novo} generation is inherently open-ended: it seeks to design both composition and structure simultaneously, often guided by desired property targets rather than stability alone. This makes \textit{de novo} approaches particularly powerful for inverse design, where the goal is to discover materials with tailored functionalities beyond those found in existing databases.

While property prediction of crystal structures is mainly achieved through predictive machine learning, structure discovery generation tasks such as CSP and \textit{de novo} generation are conventionally tackled using search and optimization methods, or, more recently, generative models.

\subsection{Design by combining property prediction with high-throughput screening and search}
\label{sec:trad_screening}

\subsubsection*{Forward prediction methods}
Property prediction for crystal structures has been traditionally achieved through quantum mechanical simulation methods such as DFT, which provides energy evaluations and property predictions from first principles. However, DFT is computationally expensive and scales poorly with system size and time scale, making exhaustive searches impractical for large and complex material spaces.

As computational power grew, predictive ML models began to complement DFT-based methods. Two (interconnected) main directions developed over the last decade, namely general-purpose property prediction models which map crystal structures to materials properties, and more specialized MLIPs which map specific geometric configurations of atoms to energies and forces. Both directions are highly related in terms of underlying ML methods, but differ in the exact task definition and in the datasets and applications. GNNs such as CGCNN~\cite{CGCNN_Xie_2018}, MEGNet~\cite{MEGNET_Chen_2019}, ALIGNN~\cite{ALIGNN_Choudhary_2021}, and co-NGN~\cite{ruff2024connectivity} have also shown strong performance in predicting crystal properties directly from structural inputs. Furthermore, LLMs such as MatBERT~\cite{trewartha2022quantifying} and LLM-Prop~\cite{LLM-Prop_NiyongaboRubungo2025} have also been explored to predict properties from text-based descriptions of crystal structures, capturing symmetries and site information that conventional models may miss, but not (yet) outperforming GNNs in prediction accuracy. 
MLIPs~\cite{deringer2019machine} are trained to predict the total energy and atomic forces of three-dimensional atom configurations of both periodic and non-periodic structures. Notable architectures include MACE~\cite{batatia2023macehigherorderequivariant,batatia2024foundationmodelatomisticmaterials}, M3GNet~\cite{m3gnet}, and CHGNet~\cite{chgnet}, which have enabled fast and accurate evaluation of large systems. 

Both types of predictive models substantially accelerate property estimation compared to quantum chemical calculations and experiments, and thus enable fast screening, geometry relaxation, and property prediction across large datasets without relying solely on expensive DFT calculations or even experiments.

\subsubsection*{High-throughput virtual screening and search-based generation}
Bridging property prediction and structure discovery, high-throughput virtual screening serves as a strategy that leverages predictive models to explore large materials libraries, typically defined in advance through combinatorial enumeration of possible compositions and structures.

Computational materials design dates back to seminal work in the 1990s on limited topology methods, such as the cluster expansion~\cite{laks1992efficient, wolverton1994cluster, ceder1998predicting}, and subsequently, with increasing computational power in the 2000s and 2010s, to screening approaches that broadened the candidate list in intelligent ways~\cite{hautier2010finding, hautier2011data}, ultimately leading, to the best of our knowledge, to the first truly novel materials discovered using computational approaches~\cite{chen2012carbonophosphates, chen2013sidorenkite}.

The introduction of ML methods allowed a further speedup: by coupling fast property predictors with systematic or active-learning-based sampling of candidate materials, high-throughput virtual screening enables the efficient identification of promising materials and establishes a first step in a materials design funnel process, prior to more costly computational evaluation, multi-scale simulations, or experimental synthesis and characterization~\cite{potyrailo2011combinatorial, curtarolo2013high}.

High-throughput virtual screening has successfully led to the discovery of promising candidates across diverse material classes. For instance, computational high-throughput screening based on DFT identified a new electrocatalyst for the hydrogen evolution reaction (HER)~\cite{greeley2006computational}. Similarly, screening of perovskite metal oxides for solar light absorption reduced a vast space of 5,400 compounds to just 15 promising candidates using electronic structure calculations~\cite{castelli2012computational}.

In some cases, computational predictions have been experimentally validated. Notable success stories include the identification of IRMOF-20, which demonstrated exceptional hydrogen storage capacities out of 5,309 metal-organic frameworks (MOFs)~\cite{ahmed2017balancing}, and the discovery of high-performance crystalline Cu–S based thermoelectric materials from the ICSD database~\cite{zhang2017screening}.

Once the materials libraries become more open and thus larger, an exhaustive enumeration and high-throughput virtual screening quickly becomes unfeasible. As a result, materials design must incorporate adaptive search and optimization approaches such as evolutionary algorithms.

CSP traditionally has been achieved experimentally by synthesizing the crystals and (single crystal) diffraction pattern characterization, which can be directly inverted to determine the structure.  
However, experimentation is time-consuming and challenging, which has led to a switch of focus towards computational approaches.
Before the rise of end-to-end generative ML models, CSP relied on high-throughput screening, heuristic search strategies, and combinatorial enumeration. These methods often incorporate ``semi-generative" principles by building structures based on evolutionary concepts or physical design rules.

Among the earliest and most widely used methods is USPEX~\cite{USPEX_GLASS2006713}, an evolutionary genetic algorithm that applies operators like heredity and mutation to evolve populations of structures toward low-energy minima, guided by DFT fitness evaluations. CALYPSO also takes an evolutionary approach founded on particle swarm optimization, using symmetry-constrained random initialization, bond characterization matrices to diversify the search, and refining candidates via DFT~\cite{CALYPSO1, CALYPSO2}. AIRSS employs a conceptually simple random search approach, generating random unit cells that respect minimal physical constraints and rely on brute-force DFT relaxations for selection~\cite{AIRSS_Pickard_2011}.
Although these methods have been highly successful in identifying both stable and metastable crystal structures across diverse material classes, they remain computationally expensive, as they rely heavily on DFT for structural relaxation. 

Alongside evolutionary and stochastic methods, rule-based structure generators have been developed for the rapid creation of specific crystal frameworks like metal-organic frameworks (MOFs) or covalent organic frameworks (COFs). ToBaCCo assembles crystalline frameworks from modular building blocks and topological templates, scaling the unit cell to fit metal clusters and organic linkers~\cite{ToBaCCo_Colon2017}. AuToGraFS follows a similar principle but includes a force field for in-situ optimization and allows for controlled introduction of defects, functional groups, or supercells~\cite{autografs}. These tools enable high-throughput generation of diverse hypothetical structures, although they typically require post-processing to assess synthesizability and stability.

Recent approaches have begun integrating predictive ML with traditional search strategies to improve efficiency. GNOME is one such hybrid system: it combines AIRSS-style random sampling with property prediction using GNNs~\cite{GNOME_Merchant2023}. GNOME applies active learning to iteratively improve the predictive model as more DFT labels are acquired, enabling efficient down-selection from millions of generated candidates. Its workflow has led to the identification of over 380{,}000 novel stable crystals and tens of thousands of materials with promising energy-related applications, now cataloged in the Energy-GNoME database~\cite{deangelis2024energygnomelivingdatabaseselected}.

\begin{figure}
    \centering
    \includegraphics[width=1.0\linewidth]{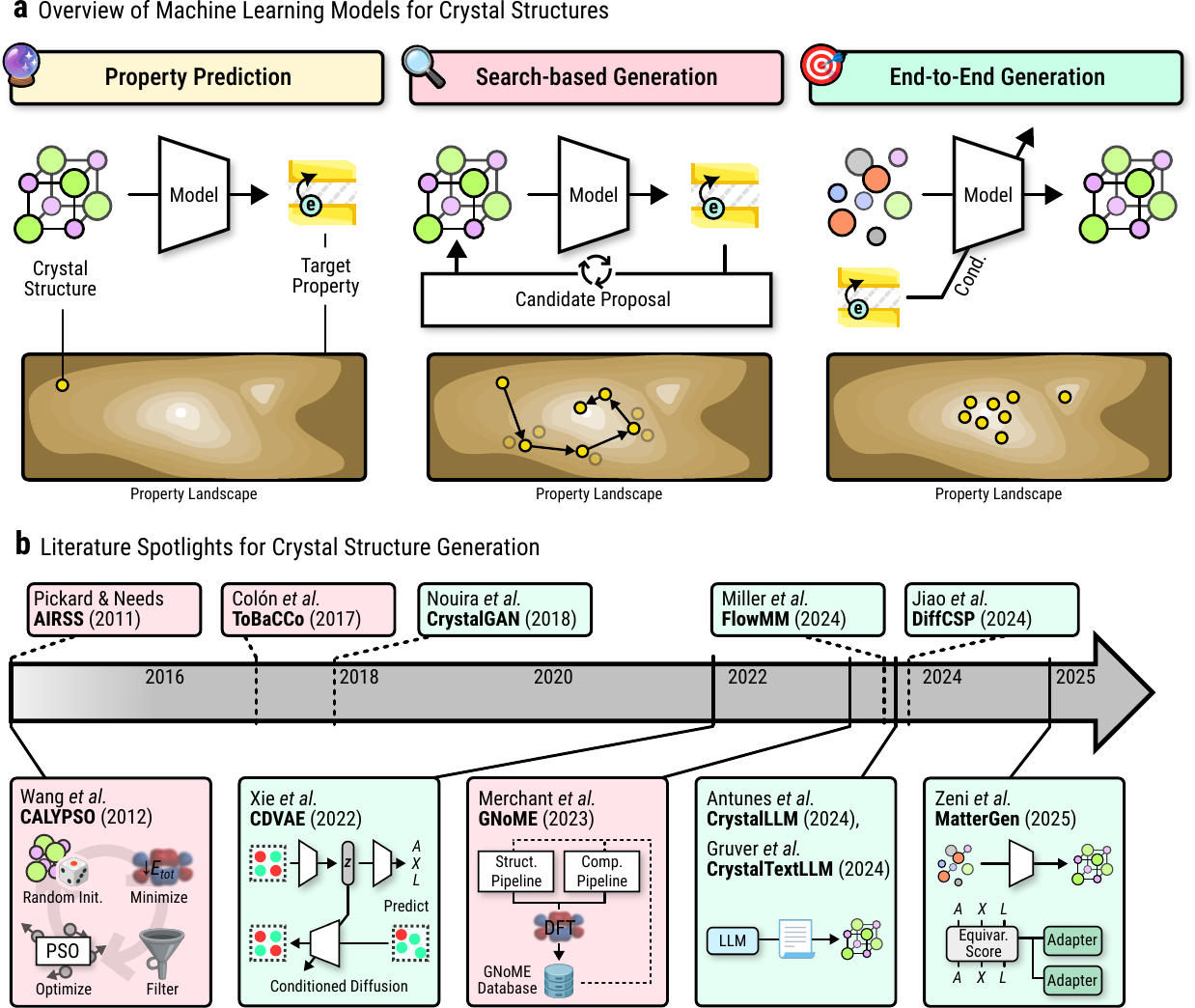}
    
    \caption{%
    \textsf{\textbf{a}} Overview of machine learning models in crystal structure generation. Left to right: Using machine learning models for simple property prediction, predictive models in conjunction with candidate proposal strategies for search-based generation, and end-to-end generation from a property conditional distribution. 
    \textsf{\textbf{b}} Literature spotlights for crystal structure generation.
    }
    \label{fig:figure_2}
\end{figure}

\begin{figure}
    \centering
    \includegraphics[width=1.0\linewidth]{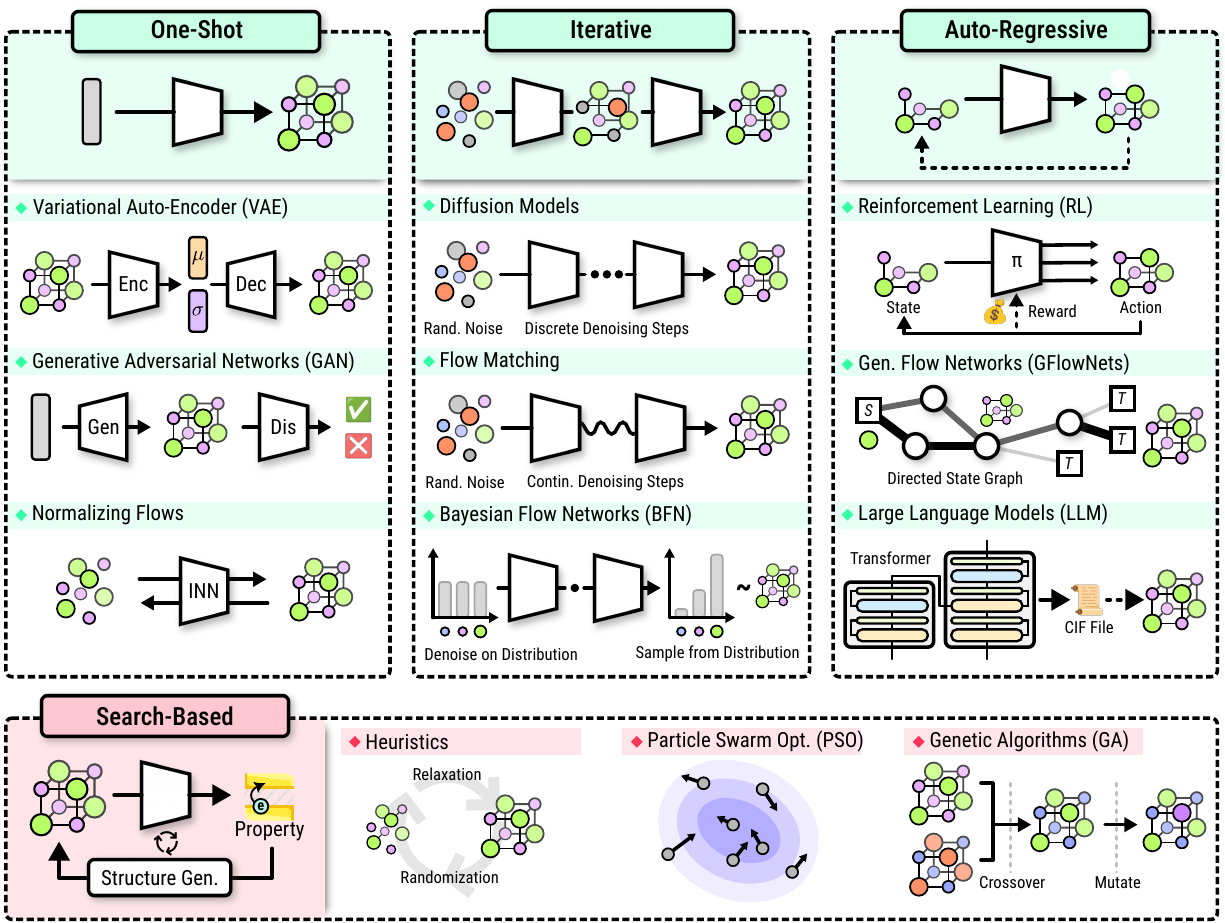}
    
    \caption{%
    Overview of different generative modeling approaches, divided into four conceptual categories. One-shot variational approaches directly generate crystal structures based on a latent space representation. Iterative methods repetitively refine the generation outcome. Auto-regressive methods iteratively construct samples from individual components of the crystal - or from tokens in the case of large language models. Search-based approaches employ property prediction models in conjunction with candidate proposal strategies, based on heuristics and/or physical intuition.
    }
    \label{fig:figure_3}
\end{figure}

Taken together, these DFT-based, ML-assisted, and heuristic search methods form the foundation of modern CSP and \textit{de novo} generation workflows. \textbf{Figure~\ref{fig:figure_2}} provides an overview of this landscape, summarizing how property prediction, search-based generation, and end-to-end generative models are connected, along with literature spotlights. These approaches illustrate a transition from purely physics-based prediction to data-augmented search and screening. These frameworks generate large volumes of plausible structures, evaluate them with ML and DFT, and refine candidates toward realistic, stable materials. This layered approach—generate, predict, validate—sets the stage for the emergence of end-to-end generative models (\textbf{Figure~\ref{fig:figure_3}}), which now aim to invert the process entirely: to learn how to directly propose structures with desired properties, a topic explored in the following sections. In Table~\ref{tab:model_comparison}, we provide a first, concise overview of major end-to-end generative model families for crystal structure design, highlighting key trade-offs in generation quality, inference speed, training stability, and likelihood estimation.
\begin{table}[h]
\centering
\resizebox{\textwidth}{!}{
\begin{tabular}{|l|l|l|l|l|}
\hline
\textbf{Model} & \textbf{Generation quality} &\textbf{Inference speed} & \textbf{Training stability} & \textbf{Likelihood estimation} \\
\hline

Variational autoencoders & Medium & \makecell[l]{Fast:\\ One-shot generation} 
& \makecell[l]{Often hard to tune} 
& Only ELBO \\
\hline

Generative adversarial networks & Medium & \makecell[l]{Fast:\\ One-shot generation} 
& \makecell[l]{Often prone to\\ minimax oscillations} 
& Not available \\
\hline

Diffusion models & \multirow{4}{*}{\makecell[l]{Competing\\for SOTA}} 
& \makecell[l]{Typically slow:\\ Many iterative steps} 
& \makecell[l]{Usually stable:\\ Score matching objectives;\\ can be sensitive to noise scheduling} 
& \makecell[l]{Possible but\\ very expensive} \\
\cline{1-1}\cline{3-5}

Flow matching &  
& \makecell[l]{Medium: Typically\\ fewer steps than diffusion} 
& \makecell[l]{Usually stable:\\ Flow-matching objectives} 
& \makecell[l]{Possible but\\ very expensive} \\
\cline{1-1}\cline{3-5}

Bayesian flow networks &  
& \makecell[l]{Medium to high: Typically\\ fewer steps than diffusion} 
& \makecell[l]{Usually stable:\\ Bayesian update framework} 
& \makecell[l]{Not reported\\ in literature} \\
\cline{1-1}\cline{3-5}

Large language models &  
& \makecell[l]{Typically slow to medium:\\ Token-by-token generation} 
& \makecell[l]{Usually stable: Standard\\ cross-entropy objective} 
& \makecell[l]{Only auto-regressive\\ likelihoods} \\
\hline

\end{tabular}
}
\caption{Qualitative comparison of model families for crystal structure generation. The entries provide approximate assessments intended to guide intuition rather than quantitative ranking. Actual performance depends strongly on representation, sampling steps, model and batch size, hardware, implementation details, and related experimental choices.}
\label{tab:model_comparison}

\end{table}

\subsection{Early end-to-end generative models}

\subsubsection*{Variational autoencoders}
Variational autoencoders (VAEs), introduced by \citeauthor{kingma2013auto}, are generative models that integrate deep neural networks with principles of variational inference~\cite{kingma2013auto}. Their generative capacity has led to successful applications in domains like image synthesis~\cite{Razavi2019VAEImage, pu2016variational}. VAEs have been adapted for complex scientific tasks, including \textit{de novo} molecular design and the generation of new crystal structures.

VAEs are composed of a probabilistic encoder, which maps inputs to a distribution in a latent space, and a decoder, which reconstructs data from its latent representation. The training objective minimizes both a reconstruction error and the Kullback-Leibler divergence, which promotes a smooth, continuous, and well-structured latent space. This probabilistic framework enables the resulting model to be highly effective at both reducing dimensionality and generating new data samples. 

Early approaches often used direct, image-like representations. For example, iMatGen represents the unit cell with voxels on a fine grid~\cite{noh_inverse_2019}.
The FTCP (Fourier-transformed crystal properties) framework refines this concept by using a Fourier-transformed representation, which introduces physical constraints to ensure the grid is mathematically invertible~\cite{ren_invertible_2022}.

Instead of relying on direct spatial grids, other VAE models operate on a more abstract level, using descriptive parameters to define a structure. As an example, PCVAE represents a crystal’s geometry using chemistry- and physics-informed descriptors such as the Bravais lattice and space group, and then decodes them to classify the space group and determine the lattice constants~\cite{liu_pcvae_2023}.

Building on the idea of a structured latent space, MagGen uses graph-theoretic analysis to quantify structural similarity, streamlining generation and enforcing design constraints~\cite{mal_maggen_2024}. Other methods focus on improving the physical realism of the generated crystals. WyCryst, for instance, produces more plausible structures by adding penalty terms to its model, ensuring the crystals adhere to both standard Euclidean symmetries and specific constraints within their allowed space groups~\cite{zhu_wycryst_2024}.

Beyond these architectural innovations, new training strategies have also emerged to guide the discovery process. The evolutionary VAE for perovskite discovery (EVAPD) uses genetic algorithms to identify and rank the most promising candidates~\cite{chenebuah_evolutionary_2023}, while TL-VAE improves outcomes through target learning~\cite{chenebuah_target-learning_2023}.

Finally, a more fundamental solution to handling complex crystal properties has been the development of graph-based VAEs. These models are naturally suited to representing structural symmetries, moving beyond both grid representations and abstract descriptors. A key example is the model developed by \citeauthor{winter_permutation-invariant_2021}, which is permutation-invariant by design, ensuring that symmetry is inherently respected throughout the generation process~\cite{winter_permutation-invariant_2021}.

\subsubsection*{Generative adversarial networks}
Generative adversarial networks (GANs), first proposed by \citeauthor{goodfellowGenerativeAdversarialNetworks2014}, are generative models that consist of two neural networks—a generator and a discriminator—engaged in a minimax game~\cite{goodfellowGenerativeAdversarialNetworks2014}. The generator aims to produce synthetic data that mimics real samples, while the discriminator learns to distinguish between real and generated data. Through this competitive interplay, both networks iteratively improve until the generator produces data that the discriminator can no longer reliably differentiate from real examples. This adversarial training allows GANs to model complex data distributions without relying on explicit probability functions. Due to their ability to produce high-quality synthetic data, GANs have achieved notable success in tasks such as image generation, style conversion, and data enhancement~\cite{gui2021review}. In addition, GANs have also been adapted for advanced scientific challenges, such as the design of novel molecules.

CrystalGAN is the first model to adapt GAN architectures to the synthesis of inorganic crystal structures~\cite{nouiraCrystalGANLearningDiscover2019}, demonstrating that generative models could learn structural chemistry rules sufficiently well to produce stable crystals beyond the prototypes included in the training data.

Building upon this foundation, later research focused on guiding the generative process toward specific material properties. ZeoGAN introduces the concept of inverse design by conditioning generation on adsorption energy grids~\cite{kimInverseDesignPorous2020}. In parallel, some models explore the advantages of a two-stage strategy: first generating plausible compositions, and then predicting their corresponding crystal structures. For example, the composition-conditioned crystal GAN (CCCGAN) generates crystals consistent with prescribed chemical formulas and validates their stability using DFT calculations~\cite{kimGenerativeAdversarialNetworks2020}, while MatGAN focuses exclusively on producing chemically reasonable and charge-balanced compositions~\cite{danGenerativeAdversarialNetworks2020}.

Efforts to extend generative capabilities further have led to CubicGAN~\cite{zhaoHighThroughputDiscoveryNovel2021}, which scales the approach to a dataset of over 370,000 ternary materials and successfully produces hundreds of new phonon-stable cubic structures. Meanwhile, the constrained crystals deep convolutional GAN (CCDCGAN) incorporates physical constraints directly into its training process by penalizing high-energy configurations~\cite{longConstrainedCrystalsDeep2021}. This integration of energy-informed objectives marked a broader trend toward property-constrained generation, ensuring that chemical realism is maintained alongside structural novelty.

More recent advances have increasingly integrated crystallographic principles into GAN architectures to enhance the physical validity of generated structures. The physics-guided crystal generative model PGCGM explicitly encodes space-group symmetries, addressing a common limitation of unconstrained models that tend to produce low-symmetry or distorted structures~\cite{zhao2023physics}. Extending this paradigm, the crystal generative Wyckoff GAN (CGWGAN)~\cite{suCGWGANCrystalGenerative2024} incorporates asymmetric units and Wyckoff positions to guarantee that generated templates satisfy both symmetry and atom-count requirements from the outset.

\subsubsection*{Reinforcement learning}
Reinforcement learning (RL) is a branch of ML where an agent learns by interacting with its environment and receiving rewards or penalties. In the context of crystal generation, RL can help navigate the vast design space by learning strategies that lead to stable structures or materials with desired properties. Over the past few years, RL applications in materials science have evolved from accelerating structure searches to directly designing crystals with specific functionalities.

Early applications of RL to crystal problems focused on accelerating CSP. \citeauthor{rlcsp} developed RL-CSP to dynamically choose moves such as atom swaps or lattice perturbations during CSP~\cite{rlcsp}. Compared to fixed, hand-tuned policies, this adaptive approach reduced the number of calculations required by up to 46 \%, showing that agents could “learn” how to reach stable structures more efficiently. Similarly, \citeauthor{meldgaardStructurePredictionSurface2020} applied deep RL to surface reconstruction, where an agent proposed atomic placements for TiO$_2$ and SnO$_2$ surfaces, guided by DFT energies~\cite{meldgaardStructurePredictionSurface2020}. This demonstrated that RL can handle not only bulk structures but also the complex problem of predicting surface configurations.

RL has also been applied in inverse design to generate stable structures with target properties. 
For example, using offline RL as a way to guide crystal design, where agents trained on past data could generate new compositions whose band gaps matched desired targets~\cite{LearningConditionalPolicies}. By learning from past data instead of online calculations, the model suggests chemically realistic structures while moving toward property goals. This showed that RL can shift from exploration to purposeful design.

Soon after, RL was extended to multi-objective optimization. RL agents were combined with predictive models to propose new inorganic materials that satisfied multiple criteria at once, such as band gap, mechanical strength, and even synthetic accessibility. By balancing these objectives, RL-based methods began to resemble practical design tools, addressing the reality that materials must perform across several metrics, not just one~\cite{karpovichDeepReinforcementLearning2024}.

Most recently, \citeauthor{MatInvent} combined RL with diffusion-based generative models for crystal structures~\cite{MatInvent}. Here, crystal generation was treated as a multi-step decision process within a diffusion model, and RL rewards guided the model toward crystal structures with target properties while maintaining diversity. This hybrid approach achieved higher success rates in goal-directed design than either RL or diffusion models alone, showing the potential of combining generative modeling with adaptive learning, as demonstrated recently in other works~\cite{mohanty2024CrysTextGenerativeAI,park2025guidinggenerativemodelsuncover}. 

While early end-to-end generative approaches, such as VAEs, GANs, and RL, proved useful in early CSP and \textit{de novo} crystal design, newer architectures, presented in the following sections, surpass them by better capturing physical symmetries and material constraints, among other reasons.
\subsection{Recent end-to-end generative models}
\subsubsection*{Diffusion models}

First introduced by \citeauthor{sohl2015deep}, diffusion-based models have rapidly become the dominant approach for image, video, and audio generation~\cite{sohl2015deep, yang2023diffusion}. More recently, diffusion models have gained increasing attention in graph-based data generation and are now considered the state-of-the-art method for generating molecular structures~\cite{alakhdar2024diffusion,wang2025diffusion}.

The essential idea behind diffusion-based models is to systematically and gradually destroy structure in a data distribution, and then to learn to reverse this process to restore the data. This framework, while allowing several formulations such as Denoising Diffusion Probabilistic Models~\cite{ho2020denoising}, Score Matching with Langevin Dynamics~\cite{song2019generative}, and the generalized Stochastic Differential Equations~\cite{song2020score}, invariantly consists of two parts: a forward diffusion process, where noise is progressively introduced into real data until it approximates a chosen prior distribution, and a reverse generation process, where a model (typically parametrized with a neural network) is trained to recover the data distribution from noise.

The application of diffusion models to crystal structure generation was first introduced by \citeauthor{xie2021crystal} with their crystal diffusion variational autoencoder (CDVAE)~\cite{xie2021crystal}. This model combines a VAE with a denoising diffusion model. Specifically, CDVAE uses a VAE to encode the joint distribution of composition, lattice, and the number of atoms. A diffusion-based model then learns the joint distribution of atomic coordinates and types, conditioned on the latent code from the VAE. CDVAE explicitly encodes the interactions across periodic boundaries, ensuring it respects permutation, translation, rotation, and periodic invariances inherent in crystal structures. Subsequent works~\cite{yeConCDVAEMethodConditional2024,luoDeepLearningGenerative2024} have extended the model by enabling the conditional generation of crystal structures with target properties.

\citeauthor{jiao2024CrystalStructurePrediction} proposed DiffCSP, which introduces several novelties into diffusion-based crystal generation~\cite{jiao2024CrystalStructurePrediction}. DiffCSP is the first generative model to use diffusion for jointly modeling atom type, lattice parameters, and atom coordinates, resulting in a more accurate representation of crystal geometry. The model operates on fractional coordinates, and it inherently encodes periodicity by employing the wrapped normal (WN) distribution~\cite{de2022riemannian}. Additionally, DiffCSP uses the Equivariant Graph Neural Network (EGNN) as a denoising network~\cite{satorras2021n}, thereby ensuring E(3) invariances for both lattice parameters and fractional coordinates. Many other frameworks in the literature have later adopted EGNN.

A significant advancement in the capabilities of diffusion-based generative models for crystals is the introduction of symmetry-aware models. Several recent models in the literature explore various approaches to ensure that the generated structures belong to specific space groups. Built on their previous work, \citeauthor{jiao2024SpaceGroupConstrained} introduced DiffCSP++, which conditions its diffusion process on space groups and leverages predefined structural templates extracted from the training data~\cite{jiao2024SpaceGroupConstrained}. This method ensures high-fidelity symmetry but potentially limits the discovery of novel structures beyond its training data. \citeauthor{levy2025SymmCDSymmetryPreservingCrystal} offered an alternative approach~\cite{levy2025SymmCDSymmetryPreservingCrystal}. SymmCD generates crystal structures by jointly denoising a compact representation of the asymmetric unit—comprising lattice parameters, atom types, fractional coordinates, and site symmetries—while explicitly preserving space group constraints throughout the diffusion process. This allows for realistic and diverse space group distributions while avoiding reliance on templates. In contrast, WyckoffDiff inherently preserves space-group symmetry by directly diffusing over Wyckoff positions and site occupancies, thus encoding symmetry by construction rather than learning it~\cite{kelvinius2025WyckoffDiffGenerativeDiffusion}.

Given the flexibility and sample quality demonstrated by diffusion-based approaches to generating crystalline structures, several specialized models have been developed for specific classes of materials. Diffusion models have been successfully applied for generating, among others, coarse-grained MOFs~\cite{fu2023mofdiff}, new families of hypothetical superconductors~\cite{yuan2024diffusion}, and zeolites with targeted properties~\cite{park2024inverse}.

Introduced by~\citeauthor{zeni2025GenerativeModelInorganic}, MatterGen exemplifies the practical capabilities of diffusion models for crystal generation~\cite{zeni2025GenerativeModelInorganic}. It applies diffusion jointly to the entire crystal representation, including lattice parameters, fractional coordinates, and atom types. MatterGen also incorporates adapter modules for fine-tuning the base model to achieve desired chemical compositions, symmetries, and scalar properties, demonstrating its effectiveness across magnetic, electronic, and mechanical domains. The model was trained on the large-scale Alexandria dataset~\cite{ghahremanpour2018alexandria}, and its generated structures have undergone partial experimental validation, with usable code and pre-trained weights made publicly available.

Finally, Chemeleon represents a hybrid diffusion framework that combines text-conditioned generative modeling with contrastive representation learning, mapping textual and crystal graph embeddings into a shared latent space~\cite{park2025exploration}.
Building on this, a more recent hybrid approach extends Chemeleon’s text-conditioned diffusion backbone by incorporating a reinforcement learning (RL) module that guides latent diffusion sampling through multi-objective rewards, thereby coupling diffusion-based generation with RL-driven exploration of the latent space~\cite{park2025guidinggenerativemodelsuncover}.

\subsubsection*{Normalizing flows}
\label{sec:flow-based_methods}

Normalizing flows relate a complex target distribution to a simpler base distribution (often Gaussian or uniform) via an invertible function.

Discrete normalizing flows parametrize an invertible neural network, balancing expressiveness with the need for tractable Jacobian determinants.
The Jacobian determinant is used to calculate the likelihood of each sample using the change-of-variables formula, which allows maximum-likelihood training ~\cite{nice,normFlowKoby,normFlowPapamakarios,realNVP,neuralSplineFlows}.

In the realm of crystal structure modeling, discrete normalizing flows are often employed as Boltzmann generators that model the full Boltzmann distribution for a given composition~\cite{boltzmannGenerator}. This is in line with the CSP task, though instead of only generating stable structures for a given composition, the full thermodynamic Boltzmann distribution is modeled. This can be used to determine (formation) free energies of crystal structures, which typically requires access to exact likelihoods.
Since most other types of generative models lack straightforward access to likelihoods, discrete normalizing flows are a natural choice.

\citeauthor{wirnsbergerTargetedFreeEnergy2020a} used normalizing flows to calculate free energy differences between two thermodynamic states~\cite{wirnsbergerTargetedFreeEnergy2020a}. The flow acts as an invertible mapping between the states to increase overlap compared to traditional free energy perturbation methods, yielding improved free energy estimates.
Furthermore, \citeauthor{wirnsbergerNormalizingFlowsAtomic2022} trained a normalizing flow to model the Boltzmann distribution of Lennard‑Jones crystals and cubic ice I to obtain absolute Helmholtz free energy estimates~\cite{wirnsbergerNormalizingFlowsAtomic2022}.
Similarly, \citeauthor{ahmadFreeEnergyCalculation2022} applied normalizing flows to compute Gibbs free energies of perfect diamond‑cubic Si crystals and monovacancy defect formation energies~\cite{ahmadFreeEnergyCalculation2022}. 
\citeauthor{kohlerRigidBodyFlows2023} introduced rigid‑body flows to model positions and orientations of molecules in molecular crystals~\cite{kohlerRigidBodyFlows2023}. 
They used this approach to calculate the free energy difference between two thermodynamic states of the hydrogen-ordered crystal phase of water. \citeauthor{schebekEfficientMappingPhase2024b} developed conditional normalizing flows to map a single reference equilibrium distribution across a range of temperatures and pressures for accurate phase diagram prediction~\cite{schebekEfficientMappingPhase2024b}. Applied to a Lennard-Jones face-centered cubic crystal, their model accurately predicted the solid–liquid coexistence line while reducing the required sampling compared to traditional methods. 

While discrete normalizing flows transform a prior distribution to a target using a finite number of invertible neural network layers,
continuous normalizing flows are the continuous generalization thereof~\cite{chenNeuralOrdinaryDifferential2018,grathwohlFFJORDFreeFormContinuous2018a}. Here, samples from a prior distribution are transformed by integrating an ordinary differential equation, moving the samples along a time-dependent vector field to match the target distribution.
Analogous to discrete normalizing flows, the likelihood of a sample can be obtained by a continuous generalization of the change-of-variables formula. However, this requires the evaluation of the vector field divergence in each integration step, which is often prohibitively expensive. This makes them less suitable for calculating free energies compared to discrete normalizing flows.

However, continuous normalizing flows are typically more expressive compared to discrete normalizing flows, as they do not suffer from the same architectural constraints.
The conditional flow matching training objective allows efficient simulation-free training of continuous normalizing flows~\cite{lipmanFlowMatchingGenerative2022,liuFlowStraightFast2022}, similar to score matching in diffusion models. This makes continuous normalizing flows a promising competitor to diffusion models, which is attributed to the fewer integration steps they often require during sampling due to more efficient trajectories~\cite{lipmanFlowMatchingGenerative2022}.
Recently, Riemannian flow matching was introduced, which extends flow matching to general manifolds~\cite{chenFlowMatchingGeneral2023}.
Based on this approach, \citeauthor{millerFlowMMGeneratingMaterials2024a} introduced FlowMM, a crystal generation framework that uses Riemannian Flow Matching to handle the periodic boundaries inherent in crystal modeling~\cite{millerFlowMMGeneratingMaterials2024a}. Similarly,
\citeauthor{kimMOFFlowFlowMatching2025} used Riemannian flow matching to model MOFs, treating the metal nodes and organic linkers as rigid bodies to reduce the dimensionality of the problem~\cite{kimMOFFlowFlowMatching2025}.
\subsubsection*{Generative flow networks}

Generative flow networks (GFlowNets) represent a class of generative models that learn a stochastic policy to sequentially construct compositional objects, such as crystal structures, by traversing a directed acyclic graph from an initial state to a terminal state~\cite{bengioFlowNetworkBased2021a,bengioGFlowNetFoundations2023}. Unlike traditional generative models that aim to maximize likelihood or minimize reconstruction error, GFlowNets are designed to sample diverse objects from a set of terminal states with probabilities proportional to a user-defined positive reward function. This proportionality-to-reward sampling enables efficient exploration of vast search spaces, facilitating the discovery of multiple high-reward candidates rather than converging on a single optimum. Training objectives such as trajectory balance are commonly used to ensure consistency between the forward generation policy and an auxiliary backward policy~\cite{bengioFlowNetworkBased2021a,bengioGFlowNetFoundations2023}.

Crystal-GFN introduced by \citeauthor{hernandez-garciaCrystalGFNSamplingCrystals2023} is a foundational GFlowNet model for inorganic crystal structures~\cite{hernandez-garciaCrystalGFNSamplingCrystals2023}. It employs a sequential generation scheme, first selecting the space group, then the chemical composition, and finally the lattice parameters of a unit cell. This approach allows for the systematic incorporation of fundamental crystallographic constraints, such as charge neutrality and compatibility between composition, space group, and lattice geometry, at each step. The model is guided by a reward signal from a proxy model predicting formation energy, enabling it to sample structures proportional to their predicted stability.

Building on this, \citeauthor{nguyenEfficientSymmetryAwareMaterials2024} proposed a symmetry-aware hierarchical architecture for flow-based traversal (SHAFT)~\cite{nguyenEfficientSymmetryAwareMaterials2024}. SHAFT refines the hierarchical generation process by initially selecting a high-level crystallographic prototype (space group), then determining consistent lattice parameters, and finally adding atoms into the unit cell while adhering to these constraints. A key innovation is the explicit exploitation of crystal symmetries; once a space group is chosen, symmetric atomic positions are automatically populated, reducing the search space complexity. Guided by a physics-informed reward function incorporating formation energy and symmetry preferences, SHAFT has shown superior performance in generating valid, diverse, and stable crystal structures compared to non-hierarchical GFlowNets and diffusion-based VAEs.

In another work, \citeauthor{cipciganDiscoveryNovelReticular2024} proposed MatGfn, which has been used to design reticular materials like MOFs and COFs for carbon dioxide capture, by building string sequences into CIFs based on a gravimetric surface area reward~\cite{cipciganDiscoveryNovelReticular2024}.
\subsubsection*{Bayesian flow networks}
In 2023, \citeauthor{graves2025bayesianflownetworks} introduced Bayesian flow networks (BFNs), a novel class of generative neural networks~\cite{graves2025bayesianflownetworks}. Similar to diffusion models, BFNs generate new samples through an iterative process. However, they differ by explicitly modeling the underlying data distributions rather than transforming or denoising the data directly. This approach enables a fully continuous and differentiable generation process, even for discrete data. Generation starts from independent, simple prior distributions (\textit{e.g.}, standard normal for continuous data or uniform categorical for discrete data), which are iteratively updated via Bayesian inference using noisy samples and refined through a neural network that captures interdependencies among variables. The final sample is drawn from the resulting distribution after the last step. By updating the parameters of the underlying distributions instead of individual data points, BFNs generate high-quality samples in just a few iterations, offering a more efficient alternative to traditional diffusion and flow-based models~\cite{wu2025PeriodicBayesianFlow,ruple2025symmetryawarebayesianflownetworks}.

\citeauthor{wu2025PeriodicBayesianFlow} introduced BFNs for crystal structures by extending the Bayesian flow to non-Euclidean manifolds to account for the periodic nature of fractional coordinates~\cite{wu2025PeriodicBayesianFlow}. \citeauthor{ruple2025symmetryawarebayesianflownetworks} implemented a symmetry-aware BFN that operates on the asymmetric unit representation, ensuring efficient generation while respecting crystal symmetries~\cite{ruple2025symmetryawarebayesianflownetworks}.
\subsubsection*{Transformer models and large language models (LLMs)}
Transformer models, originally introduced in the context of natural language processing, are based on self-attention mechanisms that allow them to learn complex relationships within sequential data. An LLM is based on variants of the transformer model~\cite{vaswani2017attention}, trained on vast corpora of text data to learn the statistical structure of language. LLMs and transformer models can be adapted to material science, either by fine-tuning general-purpose LLMs on text representations of crystal structures or by using transformers with tokenization schemes specific to crystal structure text representations. The latter approach has shown greater promise in current benchmarks.
In this context, tokenization refers to converting structured crystal data (such as CIF files) into a linear sequence of discrete symbols (tokens) that a language model can process. Standard NLP tokenizers (\textit{e.g.}, byte-pair encoding) are often ill-suited to scientific formats, prompting the development of domain-specific tokenizers that better reflect the grammar and symmetries of crystallographic data. This allows full flexibility to capture domain-specific structure and symmetries (\textit{e.g.}, SE(3) invariance), but requires large datasets and significant computational resources.

One of the earliest such methods, XYZTransformer~\cite{flam-shepherd2023LanguageModelsCan}, adopts a GPT-style architecture~\cite{radfordImprovingLanguageUnderstanding}, leveraging tokenization derived from compacted CIF files and incorporating data augmentation to learn crystallographic invariances. Similarly, CrystaLLM~\cite{antunes2024CrystalStructureGeneration} builds on the nanoGPT architecture~\cite{nanogpt}, employing CIF-specific tokenization and unsupervised pretraining on 2.3 million randomly sampled CIF files. Mat2Seq, also builds on the nanoGPT architecture like CrystaLLM, introduces SE(3)-invariant sequence representations to capture the rotational and translational symmetries inherent in crystal structures~\cite{yan2025InvariantTokenizationCrystalline}.
CrystalFormer~\cite{cao2024SpaceGroupInformed} and WyFormer~\cite{kazeev2025WyckoffTransformerGeneration} adopt autoregressive Transformer architectures that explicitly incorporate crystallographic symmetries by conditioning on the space group and applying tokenization schemes based on Wyckoff positions, which also contributes to improved inference speed.

On the other hand, fine-tuning refers to adapting an already pretrained LLM, such as LLaMA or Mistral, to a specific domain by training it further on domain-specific data (\textit{e.g.}, CIF files). This is more efficient, as the model retains useful general representations learned from broad text corpora, and only needs to adjust its weights for the new task. Techniques such as LoRA (Low-Rank Adaptation) and QLoRA (Quantized LoRA) enable parameter-efficient fine-tuning by modifying only a small subset of the model’s layers. This reduces computational cost while retaining performance.
Crystal-text-LLM~\cite{gruver2024FineTunedLanguageModels} marks the first model to fine-tune a LLaMA-2~\cite{llama2} LLM on CIF files using LoRA. Building on this, FlowLLM~\cite{sriram2024FlowLLMFlowMatching} and CrysLLMGen~\cite{khastagir2025llmmeetsdiffusionhybrid} combine generation via the fine-tuned LLM with subsequent structural refinement through FlowMM~\cite{millerFlowMMGeneratingMaterials2024a} and DiffCSP~\cite{jiao2024CrystalStructurePrediction} optimization, respectively.
In a similar vein, \citeauthor{choudhary2024AtomGPTAtomisticGenerative} fine-tuned the Mistral-AI model~\cite{jiang2023Mistral7B} on CIF files using LoRA, while \citeauthor{mohanty2024CrysTextGenerativeAI} fine-tuned the Llama-3.1-8B model with QLoRA to generate structure descriptions~\cite{choudhary2024AtomGPTAtomisticGenerative,mohanty2024CrysTextGenerativeAI}.\\

\section{Practical considerations}
\subsection{Constraining and conditioning models} 

\begin{figure}
    \centering
    \includegraphics[width=1.0\linewidth]{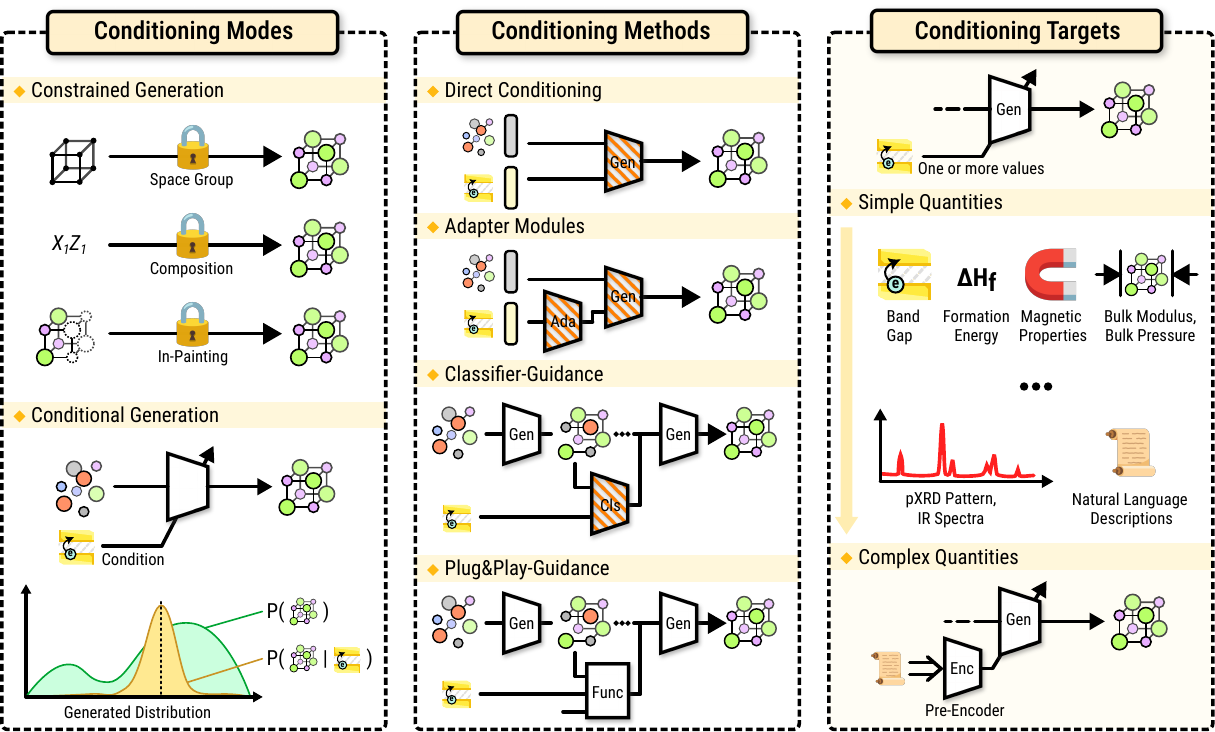}
    
    \caption{%
    Overview of targeted crystal structure generation. Left: Difference between constrained generation and property-conditioned generation. Middle: Architecture integrations of different conditioning methods. The hatched symbols represent components that are coupled to the specific property and which have to be retrained when switching properties. Right: example properties that can be used as conditioning targets for crystal structure generation. Valid conditioning targets range from simple, single-value properties, such as the band gap, to complex properties, including natural language property descriptions.
    }
    \label{fig:figure_4}
\end{figure}

While unconstrained \textit{de novo} generation of crystal structures is an interesting challenge, many practical materials discovery tasks demand a more targeted approach: generating structures that satisfy predefined constraints or exhibit specific properties. Targeted generation is central to inverse materials design, where the goal is not merely to explore structural space but to find candidates optimized for function and feasibility. This section outlines possible scenarios in targeted generation, also illustrated in \textbf{Figure~\ref{fig:figure_4}}.

A first step toward targeted generation is to incorporate structural or compositional constraints into the generation process.
Several recent approaches have demonstrated that it is possible to generate crystal structures constrained to a desired space group by leveraging its symmetry operations during generation and explicitly conditioning the model on the target group~\cite{jiao2024SpaceGroupConstrained,levy2025SymmCDSymmetryPreservingCrystal,ruple2025symmetryawarebayesianflownetworks}. This is valuable in real-world settings, as many families of materials, such as perovskites or zeolites, predominantly adopt specific space groups.

Another important constraint is chemical composition. By fixing the atom types, the task reduces from \textit{de novo} generation to CSP. For diffusion models and flow matching, fixing the atom types during generation is straightforward, enabling the model to sample from the distribution of stable configurations for a given composition. Also, inpainting generation based on known host structures is possible~\cite{zhongPracticalApproachesCrystal2025}.

Lastly, given applications define desired target properties, which can be achieved by conditioning models to generate crystal structures that fulfill them, which will be discussed later.

\subsubsection*{Symmetry-constrained generation} \label{sec:discussion:symmetries}

%

Symmetries play a fundamental role in the modeling of crystal structures (Figure~\ref{fig:figure_5}). Unlike general molecular systems, crystals must satisfy not only the Euclidean invariances of three-dimensional space, \textit{i.e.}, translations, rotations, and reflections, collectively described by the E(3) group, but also the more restrictive symmetries defined by their specific space groups. Early generative approaches typically only accounted for E(3) invariance but neglected space group symmetries~\cite{xie2021crystal,jiao2024CrystalStructurePrediction,millerFlowMMGeneratingMaterials2024a}. As a result, these models frequently generated crystals with low symmetry and made conditioning on a target space group challenging or infeasible~\cite{levy2025SymmCDSymmetryPreservingCrystal}. To address this limitation, more recent models explicitly incorporate space group information into the generation process. These methods either constrain the placement of atoms to valid Wyckoff positions consistent with the desired space group~\cite{jiao2024SpaceGroupConstrained}, or focus on generating only the asymmetric unit, which is then expanded into the full periodic crystal during post-processing~\cite{levy2025SymmCDSymmetryPreservingCrystal,ruple2025symmetryawarebayesianflownetworks,cao2024SpaceGroupInformed,kazeev2025WyckoffTransformerGeneration}. This inductive bias improves the physical plausibility of the generated crystal structures and enables the restriction of the generation to specific space groups. Furthermore, models that generate only the asymmetric unit are more computationally efficient and scalable for both iterative and autoregressive architectures, as they operate on a reduced representation while ensuring full structural symmetry.
\subsubsection*{Property-conditioned generation}\label{sec:discussion:conditioning}
In practical, application-driven materials discovery, structural or chemical constraints are not sufficient. Often, the goal is to find materials that possess target properties tailored for specific applications.

Enforcing property constraints during generation is difficult due to the complex relationship between structure and property. Instead, conditional generation is generally approached via soft conditioning, where the model learns to approximate the conditional distribution of structures given a target label. This is typically done by injecting a learned representation of one or more target properties into the architecture~\cite{yeConCDVAEMethodConditional2024,zeni2025GenerativeModelInorganic,ruple2025symmetryawarebayesianflownetworks}.
Classifier-free guidance can be used to tune the influence of the target condition~\cite{hoClassifierFreeDiffusionGuidance2022,zhengGuidedFlowsGenerative2023}.

Directly conditioning a generative model requires retraining from scratch for every new target or property of interest, limiting the method’s generality and practicality.
To address this, MatterGen uses lightweight adapter modules that inject conditioning information additively into the layers of a pre-trained generative model~\cite{zeni2025GenerativeModelInorganic}. This decouples the conditioning mechanism from the core architecture and allows models to be first trained generically, then fine-tuned for specific properties with relatively little additional data or compute.

There are many properties suitable for conditional crystal generation, and the choice largely depends on the specific application.
First, scalar properties can be used for conditional crystal generation, \textit{e.g.}, electronic properties such as the band gap~\cite{yeConCDVAEMethodConditional2024,zeni2025GenerativeModelInorganic,ruple2025symmetryawarebayesianflownetworks}, physical properties such as the bulk modulus~\cite{zeni2025GenerativeModelInorganic}/bulk pressure~\cite{luoDeepLearningGenerative2024}, magnetic properties~\cite{zeni2025GenerativeModelInorganic}, or the thermodynamic stability given by the formation energy~\cite{yeConCDVAEMethodConditional2024,ruple2025symmetryawarebayesianflownetworks}. Recently, \citeauthor{parkExplorationCrystalChemical2025} introduced a text-guided crystal diffusion model, where the generation is conditioned on a text embedding based on the chemical composition and crystal system~\cite{parkExplorationCrystalChemical2025}.

Beyond generating functionalized materials with targeted properties, conditional crystal generation has also been used to support experimental analysis. Here, a model generates possible structures conditioned on (non-scalar) experimental measurements such as powder diffractograms~\cite{tsukaue2025crystal,rieselCrystalStructureDetermination2024,laiEndtoEndCrystalStructure2025,guoInitioStructureSolutions2025}, XANES spectra~\cite{kwonSpectroscopyGuidedDiscoveryThreeDimensional2023}, or IR spectra~\cite{chandankanakalaSpectraStructureContrastive2024}.

From a practical perspective, an interesting approach is to use guided generation without fine-tuning, where a powerful, pre-trained generative model can be guided at inference time. This plug-and-play guidance paradigm greatly expands accessibility, allowing domain experts to specify property targets without retraining or modifying the generative backbone. The first work in this direction has been performed~\cite{yuan2024diffusion,cao2024SpaceGroupInformed}. For example, \citeauthor{yuan2024diffusion} guide the generation of superconductors based on reference compounds during inference to discover new families of hypothetical superconductors~\cite{yuan2024diffusion}.

In the broader context of materials discovery, targeted generation promises to transform computational design workflows. 
Traditional screening approaches typically use a broad proposal distribution that is subsequently filtered in multiple stages. Conditional generative models are promising candidates to accelerate this traditional screening approach, generating materials closer to the requested target and requiring less filtering. Continued progress will require not only architectural innovations but also robust benchmarks, property-aware datasets, and a deeper theoretical understanding of the trade-offs between control, diversity, and fidelity in (conditional) generative modeling.
\subsection{Usability: software and code availability}

Access to open-source, ready-to-use software is crucial for researchers looking to apply generative models in practice. To provide a practical starting point, we grouped a selection of implementations that we consider to be scientifically relevant and representative of recent advances in the field. This selection is not complete and balances multiple criteria beyond usual benchmark metrics, including code documentation and usability.
\textbf{Table~\ref{tab:software}} summarizes this selection of recent generative models for materials, including their associated datasets, code availability, checkpoints, and conditioning options.

\begin{table}[h]
\resizebox{\textwidth}{!}{
\begin{tabular}{|l|l|l|l|l|l|}
\hline
\textbf{Model} & \textbf{Method}&\textbf{Datasets} & \textbf{Code} & \textbf{Checkpoints} & \textbf{Conditioning} \\
\hline
MatterGen \cite{zeni2025GenerativeModelInorganic} &Diffusion& \makecell[l]{
Alex-MP-20,\\ MP-20} &
\href{https://github.com/microsoft/mattergen}{GitHub} & 
\href{https://huggingface.co/microsoft/mattergen}{HuggingFace} & 
\makecell[l]{Composition, Space group, \\ Band gap, Bulk modulus, \\ Magnetic density, HHI score, \\ Energy above hull} \\

\hline
DiffCSP \cite{jiao2024CrystalStructurePrediction} & Diffusion & \makecell[l]{
Carbon, Perov,\\ MP-20} &
\href{https://github.com/jiaor17/DiffCSP}{GitHub} & 
\href{https://drive.google.com/drive/folders/11WOc9lTZN4hkIY7SKLCIrbsTMGy9TsoW}{Google Drive} & Composition, Formation Energy\\

\hline
DiffCSP++ \cite{jiao2024SpaceGroupConstrained} & Diffusion & \makecell[l]{
Carbon, Perov,\\ MP-20, MPTS-52} &
\href{https://github.com/jiaor17/DiffCSP-PP}{GitHub} & 
\href{https://drive.google.com/drive/folders/1FQ_b6CE09KtyGaU_r6uO8_I5JhrQmUFB}{Google Drive} & Space group constrained. Composition\\

\hline

CrysBFN \cite{wu2025PeriodicBayesianFlow} & BFN & \makecell[l]{
Carbon, Perov,\\ MP-20, MPTS-52} & 
\href{https://github.com/wu-han-lin/CrysBFN}{GitHub} & 
\href{https://drive.google.com/drive/folders/1W5kGiZYFRJZiyKyTwCdcPk9lbjTsTCO-}{Google Drive} & Composition\\

\hline
CrystalFormer \cite{cao2024SpaceGroupInformed} & Transformer & \makecell[l]{
Alex-20,\\ MP-20} &
\href{https://github.com/deepmodeling/CrystalFormer}{GitHub} & \href{https://huggingface.co/zdcao/CrystalFormer}{HuggingFace} & \makecell[l]{Space group constrained. \\Dielectric FoM, Energy above hull \\(with reinforcement fine-tuning)} \\

\hline
\end{tabular}

}
\caption{Overview of selected repositories of generative models for materials. Repositories were selected based on competitive (near state-of-the-art) performance and the availability of well-documented, reproducible code and pretrained checkpoints.}
\label{tab:software}
\end{table}

\subsection{Speed and computational cost}
Inference speed is a critical consideration in the practical deployment of generative models for crystal structure generation. A variation in inference time exists among methods, largely due to differences in architectural complexity, data representation, and sampling requirements~\cite{ruple2025symmetryawarebayesianflownetworks}. Exact inference runtimes depend heavily on hardware configurations, so comparisons between different methods are limited to estimates at the level of orders of magnitude. For iterative models, the number of sampling steps is the primary determinant of runtime, since each step involves one or two computationally expensive neural network evaluations. In many reported settings, diffusion-based generative models exhibit long inference times because achieving state-of-the-art performance often involves hundreds to thousands of iterative denoising steps during sampling~\cite{jiao2024CrystalStructurePrediction, levy2025SymmCDSymmetryPreservingCrystal}. Fine-tuned LLMs also operate at a high computational cost, with inference speeds varying over multiple orders of magnitude depending on model size. This variation reflects a tradeoff between computational efficiency and the quality of the generated structures~\cite{gruver2024FineTunedLanguageModels}, where the largest LLMs approach inference times comparable to those of diffusion models. Flow-based models operate at a similar speed to the smallest fine-tuned LLMs, as they generally require fewer sampling steps than diffusion models for generation, typically on the order of a few hundred steps~\cite{millerFlowMMGeneratingMaterials2024a}. Smaller transformer-based architectures employing symmetry-aware tokenization schemes achieve inference speeds approximately an order of magnitude faster than the smallest fine-tuned LLMs~\cite{cao2024SpaceGroupInformed}.
BFNs improve upon the number of required sampling steps, typically using fewer than 100 steps, reducing inference time by yet another order of magnitude compared to the transformer models. With the experimental setup reported for SymmBFN~\cite{ruple2025symmetryawarebayesianflownetworks}, using 100 sampling steps corresponds to generating a crystal structure in a few milliseconds on NVIDIA A100 GPUs.
\subsection{Metrics and theoretical evaluation of the models}
In this section, we review the computational metrics used to benchmark generative models in CSP and \textit{de novo} generation. These metrics, originally developed from an ML perspective, facilitate model comparison but only act as indirect proxies for the actual quality and validity of the generated crystal structures.

\subsubsection*{Crystal structure prediction}
In the CSP task, the model is evaluated on its ability to reconstruct the original crystal structures in the test set using only the atomic species as input. The evaluation metrics, first introduced by \textcite{xie2021crystal}, are computed using the \texttt{StructureMatcher} class from the \texttt{pymatgen} library~\cite{pymatgen}. The \texttt{StructureMatcher} defines a match by aligning two structures, comparing all valid lattice transformations and atomic arrangements within specified tolerances. The match rate is defined as the fraction of target structures for which the model generates at least one matching candidate within a fixed number of attempts. Additionally, for all matched structures, the average root-mean-square displacement between corresponding atomic sites is reported to quantify the structural similarity between matched pairs. \\
However, these metrics have important limitations. Whether a generated structure is classified as a match depends strongly on the chosen tolerance parameters, which can vary between studies and substantially affect reported match rates. Additionally, this evaluation framework considers only exact or near-exact reconstructions of the test structures, potentially disregarding novel but thermodynamically stable configurations that share the same composition yet differ structurally from the reference.
To address this limitation, it may be beneficial to additionally assess the thermodynamic stability of the generated structures, providing a more comprehensive evaluation that captures both reconstruction accuracy and the feasibility of the predicted configurations. \citeauthor{martirossyan2025structurematchesdoesglitter} provided an in-depth analysis of the commonly used metrics in CSP and proposed refined adaptations to address their limitations~\cite{martirossyan2025structurematchesdoesglitter}.

\subsubsection*{\textit{De novo} generation \label{sec:metrics_denovo_generation}}
Generative models for crystal structures aim to generate materials that are (\textit{i}) physically \textbf{stable}, (\textit{ii}) \textbf{unique} in the sense that a model generates a diverse set of materials, and (\textit{iii}) \textbf{novel}, meaning that the generated materials are not part of the training data.
\\
While synthesizability can only be definitively confirmed through experimental validation, such validation is constrained by limited experimental resources. Consequently, many studies instead use thermodynamic \textbf{stability} as a proxy, typically evaluating the energy above the convex hull, a metric first adopted for generative models by \textcite{zeni2025GenerativeModelInorganic}. This has since become a standard benchmark for comparing the viability of output crystal structures for generative models, although numerous complementary evaluation methods are required to assess the quality of the generated candidates (see Section~\ref{sec:post_generation}).
\\
The convex hull represents the set of lowest-energy phases at all compositions within a chemical system~\cite{bartel2022ReviewComputationalApproaches}. A material's position relative to the convex hull indicates its thermodynamic stability: if it lies on the hull, it is considered thermodynamically stable, whereas a positive energy above the hull suggests it may decompose into a combination of more stable phases. To assess the energetic stability of a generated material, its structure is first optimised, and the final energy is computed using either DFT or ML approaches, such as MLIPs. While MLIPs offer significantly faster evaluations, they typically trade off some accuracy compared to DFT. Commonly used MLIPs for evaluation, as described in earlier sections, include M3GNet~\cite{m3gnet}, CHGNet~\cite{chgnet}, MatterSim~\cite{yang2024mattersimdeeplearningatomistic}, and MACE-MP-0~\cite{batatia2024foundationmodelatomisticmaterials}. A comprehensive overview of these and other MLIPs is provided by \textcite{jacobs2025PracticalGuideMachine}, while a systematic benchmark of their performance is presented by \textcite{riebesell2024matbenchdiscoveryframework}. The stability rate denotes the proportion of generated materials whose final energies lie on ($E_{\mathrm{hull}} \leq 0.0 \text{ eV/atom}$), or near (often chosen to be around $E_{\mathrm{hull}} \leq 0.1 \text{ eV/atom}$), their respective convex hulls. These thresholds are widely used in ML studies for benchmarking generative models~\cite{jiao2024CrystalStructurePrediction,zeni2025GenerativeModelInorganic, millerFlowMMGeneratingMaterials2024a, }. However, experimental workflows often employ more stringent (lower) cutoff values, as described in Section~\ref{sec:post_generation}.
\\
Using thermodynamic stability as an evaluation metric has inherent limitations: the reference convex hulls are often incomplete, particularly for chemical systems containing a large variety of elements. This epistemic uncertainty can result in approximate stability assessments, potentially leading to false positives. Furthermore, comparing stability rates across different sets of generated structures can suffer from statistical fluctuations unless a sufficiently large number of samples, typically at least 10,000, is evaluated. Performing such extensive stability checks directly with DFT is computationally prohibitive; therefore, many recent works first employ MLIPs for pre-screening, followed by high-accuracy DFT calculations for the most promising candidates~\cite{wu2025PeriodicBayesianFlow,gruver2024FineTunedLanguageModels}. Variations in energy above hull thresholds and differences in the computational methods used to calculate the energies of predicted crystal structures often hinder the direct comparability of stability metrics across studies. Finally, a low energy above hull is necessary but not sufficient for synthesis, and therefore cannot serve as a comprehensive metric for the quality of generated structures (see Section~\ref{sec:post_generation}).
\\
In addition to energetic stability, a key objective is to produce a diverse set of novel structures that differ from those seen during training, rather than repeatedly generating identical or similar materials. Therefore, a structure is considered \textbf{unique} if no matching structure (determined, \textit{e.g.}, using the \texttt{StructureMatcher} class from the \texttt{pymatgen} library) exists within the set of generated structures. Similarly, a structure is considered \textbf{novel} if no matching structure is found in the training dataset. Both uniqueness and novelty metrics critically depend on the choice of similarity thresholds, which can significantly alter the results by either overestimating diversity or failing to distinguish genuinely distinct structures. A more detailed discussion of the limitations of novelty and uniqueness metrics is provided by \textcite{negishi2025continuousuniquenessnoveltymetrics}.
\\
A generated structure is classified as a \textit{S.U.N.} (stable, unique, and novel) material if it simultaneously satisfies the criteria for stability, uniqueness, and novelty. The S.U.N. rate represents the proportion of generated structures that qualify as S.U.N. materials. Many studies also report a generation \textit{cost} metric, either in terms of the number of function evaluations for iterative methods~\cite{millerFlowMMGeneratingMaterials2024a,wu2025PeriodicBayesianFlow}, or the wall-clock time required to generate a single (S.U.N.) structure.
\\
Earlier studies on generative models for crystal structures introduced a set of \textit{proxy metrics} to evaluate the quality of generated materials~\cite{xie2021crystal}. While widely adopted, these metrics are generally less indicative of practical synthetic accessibility than stability-based assessments. They typically include validity checks based on charge neutrality and interatomic distance thresholds, coverage recall, and precision scores computed using structural fingerprints relative to the test set, and comparisons of property distributions, such as elemental diversity and density, between generated structures and the test set. However, because these metrics are usually computed on only 1000 generated structures, they are highly sensitive to sample selection and do not allow for fair method-to-method comparisons. Reporting the variance over multiple random subsets can help capture this uncertainty and improve the interpretability of benchmarking results~\cite{levy2025SymmCDSymmetryPreservingCrystal}.


\subsection{Post generation selection workflows \label{sec:post_generation}}

\begin{figure}
    \centering
    \includegraphics[width=1.0\linewidth]{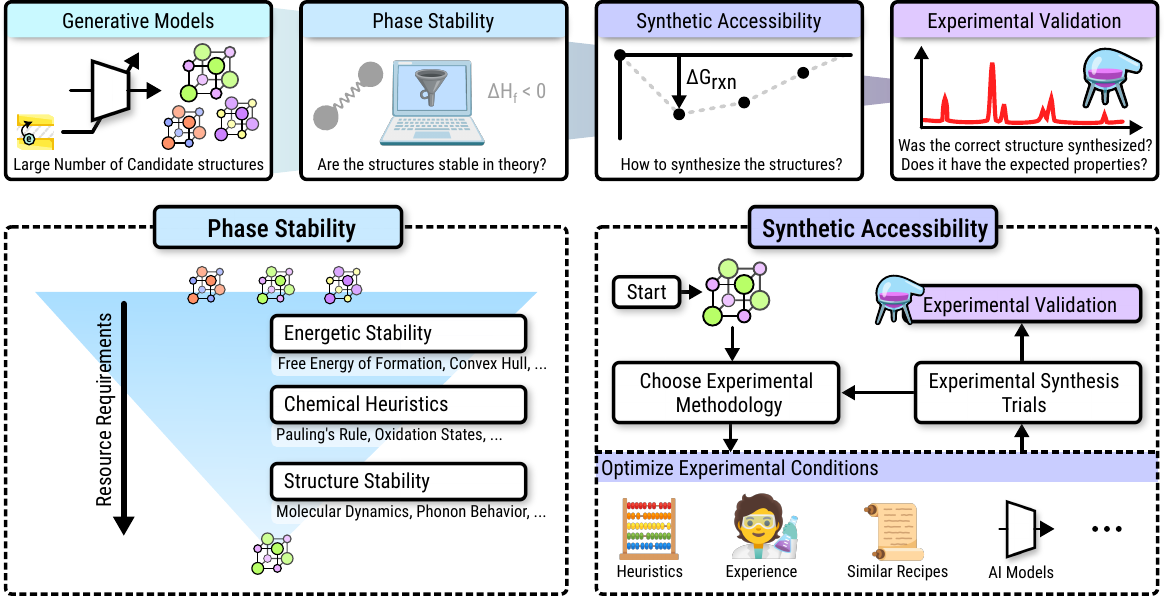}
    \caption{%
    Workflow for computationally guided experimental synthesis of generated crystal structures. Crystal structures created using generative models usually require several post-processing steps before experimental validation, which include evaluation of their phase stability (potentially in multi-step workflows depicted in the lower left), and their synthetic accessibility (lower right).
    }
    \label{fig:figure_6}
\end{figure}

Contemporary strategies for new materials development involve (generative) crystal structure prediction followed by experimental synthesis of the target material. 
However, as previously detailed, not all predicted structures may be realizable. 
Furthermore, synthesis procedure development is often cited as a bottleneck for new materials development, largely due to a lack of unified chemical synthesis theory ~\cite{Montoya2024_HardSynthesis_1, Mallouk1993_HardSynthesis_2, Jansen2018_moreModernReview}.
Commonly used trial-and-error approaches limit efficient methodology establishment and firmly place synthesis as a resource-intensive field that necessitates scrutiny for synthesis campaign planning.
Similarly, as a data-poor field, there has been very limited ML success for predictive chemical synthesis~\cite{Sun2025_SynthesisTextMining}.
Despite fragmented theory and data collections, paradigms have emerged in (high-throughput) synthesizability screening~\cite{Park2025synthesisgap}.
Shown in \textbf{Figure~\ref{fig:figure_6}}, synthesizability screening post generative crystal structure prediction generally employs filtering for target phase stability and physicality, and is followed by procedures to establish synthetic accessibility, before attempted experimental investigation.

\subsubsection*{Evaluation of target stability}
The minimum assessment for synthesizability is phase stability, agnostic of experimental method and condition. In other words, how likely is it that the generated structure is real?
In \textbf{Figure~\ref{fig:figure_6}}, we show a filtration schema for phase stability assembled from collections of target prediction and synthesizability literature.
The criteria of (\textit{i}) energetic stability, (\textit{ii}) chemical heuristics, and (\textit{iii}) structural stability are organized by approximate resource requirements, \textit{e.g.}, least to most computationally and time-intensive.
%
As previously discussed in Section~\ref{sec:metrics_denovo_generation}, many screening workflows search initially for favorable (thermodynamic) energetics, often evaluated by a negative formation energy ($\Delta H_{f}$), energy ($E_{\mathrm{hull}}$) with respect to the convex (composition-energy) hull, or appearance as the most stable phase on a phase diagram \textit{e.g.} (composition-temperature or chemical potential diagrams)~\cite{bartel2022ReviewComputationalApproaches, Ong2008_ConvexHull}.
The former parameters are easy to integrate into automatic workflows leveraging computational materials databases of bulk crystalline inorganic solids, such as the Materials Project and the Open Quantum Materials Database~\cite{Jain2013, Horton2025_MaterProj_2, Saal2013_OQMD_1, kirklin2015open}. 
Nearly all observed bulk inorganic solids have negative formation energies and are either on the hull or are within $\approx$\,100meV/atom of the convex hull, making $E_{\mathrm{hull}}$ a useful tool~\cite{Sun2016_Metastability_1, Aykol2018_Metastability_2}.
However, the threshold for metastability has been shown to be chemistry and chemical system dependent and thus prevents it from being a one-size-fits-all metric~\cite{Aykol2018_Metastability_2}.
%
Screening for chemical heuristics in proposed structures ensures that critical chemical laws known to constrain ion oxidation state, polyhedra geometries, bonding orders, and coordination numbers are followed. Historically, the evaluation of chemical rules of thumb can be challenging to enforce due to the lack of formal, rigid rules governing bonding and structure for translationally invariant bulk, crystalline solids. However, a number of studies have successfully employed specific heuristic metrics to known and candidate structures, including the Pauling rules, the Crystal Orbital Bond Index, structure-type tolerance factors, oxidation state predictors, and the \textit{mno} counting rules for intermetallic bonding~\cite{Mller2021_COBI, George2020_PaulingRules, Simonson2010_BondingScheme, Filip2018_GoldschmidtTolFac, Fu2023_OSP_1, Thway2024_OSP_2}.
Structural thermal dynamics and stability are important to understand proposed phase feasibility, and, in particular, towards assessing relative stability to a ground state atomic configuration. Structural stability can be measured through Monte Carlo (simulated annealing), molecular dynamics, and phonon calculations~\cite{Jansen2018_moreModernReview, Jansen2002_StructureEnergyLandscape, Malyi2019_EngPhDynStability, Qian2021_PhononRev, Ozoli1999_PhononScreening}, though computational and time resources are likely to be higher than for other screening metrics.
It is important to note that many high-temperature metastable polymorphs may exhibit
some lattice instability as the temperature is lowered and are therefore challenging to evaluate with high-throughput 0\,K calculations, even when phonons are considered.
%

\subsubsection*{Evaluation of synthetic accessibility}
In the absence of retrosynthesis models, route-aware synthesizability screening in advance of experimental exploration aims to ensure efficient use of time and laboratory resources when targeting a candidate structure hypothesized to be stable.
In \textbf{Figure~\ref{fig:figure_6}}, a screening workflow outlines (\textit{i}) method selection, (\textit{ii}) method feasibility, and (\textit{iii}) experimental condition prediction as the three steps that can be implemented within a predictive synthesis paradigm. 
While method selection is the first step in evaluating synthetic accessibility, this choice is often limited by individual expertise and equipment availability. Solid state (ceramic), hydrothermal, sol-gel, flux, sputtering, and deposition are among common preparative techniques~\cite{Handbook_Preparative_InOrg_Chem}.
Step \textit{ii} shows options to forecast if the considered method is appropriate for a proposed structure. Domain experts often complete this evaluation with chemical and experimental heuristics and domain knowledge: indeed, Martin Jansen wrote in 2018: ``Furthermore, the task of rationally identifying routes of syntheses to the solids predicted... has not yet been systematically tackled... this latter issue is less disturbing in practice, because a well-skilled and experienced chemist will find ways of synthesis for a material that he firmly trusts to be capable of existence"~\cite{Jansen2018_moreModernReview}.
Standardized models and descriptors to rationalize experimental methods choice are fragmented, but include creation of phase diagrams for assessment of stability at expected synthesis conditions~\cite{McDermott2023_reactionNetwork_2}, solubility tests for hydrothermal~\cite{Walters2021_Pourbaix} and sol-gel synthesis~\cite{Anantharaman2017_SolGelThermo_1, Matsoukas2015_SolGelThermo_2}, choice of reaction temperature and precursor selection~\cite{Elbashir2024_RxnTemp_1, Wenhao_inverse_hull_model, Aykol2021_PIRO}, and structure-based descriptors for crystal growth~\cite{Aykol2021_PIRO, Zur1984_LatticeMatch_1, Ding2016_LatticeMatch_2, Yang2025_TopotaticRxn_1, Liu2023_StructureforInterface}.
In the third step, suggestions for promising experimental conditions are evaluated. 
As this is an active area of research in synthesis science multiple models, programs, and theories have been proposed for different synthesis methods, including the use of simple reaction energies, more complex reaction networks ranking precursor systems by a cost function~\cite{McDermott2023_reactionNetwork_2, McDermott2021_reactionNetwork_1}; assessing number of competing phases~\cite{Aykol2021_PIRO}; phase diagrams such as Pourbaix diagrams to understand precursor solute concentration, temperature and mineralizer pH for hydrothermal and flux based methods~\cite{Walters2021_Pourbaix}; the substrate-target lattice match for epitaxial growth~\cite{Zur1984_LatticeMatch_1, Ding2016_LatticeMatch_2}; and data-driven and machine learned models~\cite{Chen2023_MLSynthModel_1, Hiszpanski2020_MLSynthModel_2, He2023_MLSynthModel_3, Boiko2023_MLSynthModel_4, Zhang2025_MLSynthModel_5}. 
We guide interested readers to an excellent review for current approaches in models and programs for predictive solid state synthesis for battery materials~\cite{Szymanski2024_CompGuidSynth}.
Experimental investigation following suggested recipe(s) can be optimized via chemical intuition, exhaustive enumeration, or active learning approaches (\textit{i.e.}, Bayesian optimization)~\cite{Shields2021_BayOptSyn_1}.

The lack of a theory of synthesizability is particularly challenging for generative structure models, as synthesis feasibility cannot be included as a constraint during the generation, but can at best be evaluated afterwards through explicit calculations of convex hulls and phase diagrams.  In principle, requirements for the topology of the convex hull near the generated compound could be included during structure generation, though this is complex due to the non-local nature of hull energy, depending on energies at many other compositions, and has therefore never been demonstrated as a feasible constraint in generative structure generation.

Despite the use of phase stability and route-aware synthetic accessibility screening, successful experimental realization is not assured~\cite{Montoya2024_HardSynthesis_1}; exhausting enumerated synthesis conditions, sometimes for multiple synthetic techniques, is likely required for confidence in inaccessibility.

\subsection{Emerging topics}

\subsubsection*{Engineering defects and disorder}
While most of the previous generative approaches focus on ideal crystalline materials, real solids can contain defects and disorder due to, \textit{e.g.}, vacancies, substitutions, and dislocations.
As these features can affect the final properties of the crystals, accounting for them is essential to bridge the gap between theoretical predictions and experimentally realizable materials.
ML and generative models for engineering defects and incorporating disorder represent an important design opportunity for broad classes of functional materials, such as 2D materials, ferroelectrics, and small-gap semiconductors, wherein computational crystal structure generation was previously tedious and incomplete.
Recent studies have begun addressing this challenge by developing models and frameworks capable of explicitly representing and learning from non-ideal structures.
It is worth noting that, to date, most approaches in the literature addressing defects and disorder rely on conventional ML models with the ability to complement generative frameworks through screening, prediction, or post-processing. Only a limited number of works, to the best of our knowledge, directly incorporate defects or disorder within the structure generation process itself.

\citeauthor{Yang2025_DefectEng_1} introduced DefiNet, a defect-informed EGNN designed to capture the local geometries and interactions in point-defect structures~\cite{Yang2025_DefectEng_1}.
A key distinction of this approach relative to conventional GNNs is the use of an explicit defect representation rather than relying on defects being implicitly inferred from atomic geometry. Each node in the graph carries additional markers that encode whether the atom corresponds to a pristine lattice site, a substitutional defect, or a vacancy, making defect identity directly available during message passing. This defect-aware formulation allows the network to learn how different defect types locally perturb atomic interactions and structural relaxations.
\citeauthor{Frey2020_DefectEng_2} applied deep transfer learning to predict defect properties in 2D materials, identifying numerous dopant–defect combinations with potential for quantum emission and neuromorphic computing~\cite{Frey2020_DefectEng_2}.
Defect effects are incorporated through engineered feature vectors that explicitly encode local chemical and structural deviations from the pristine lattice, enabling a classifier to predict defect occurrence and a regressor to estimate defect formation energies.
Mosquera-Lois et al. used a machine-learning surrogate model to accelerate defect structure relaxation, successfully predicting stable reconstructions for unseen materials and significantly reducing DFT~\cite{MosqueraLois2024_DefectEng_3}.
This framework focuses on post-generation structural refinement around defects, rather than producing full atomic configurations from scratch.
Beyond point defects, \citeauthor{Jakob2025_DefectEng_5} introduced ML classifiers to predict the likelihood of crystallographic disorder in inorganic compounds, providing a path toward disorder-aware materials discovery workflows that better mirror experimental realities~\cite{Jakob2025_DefectEng_5}.
\citeauthor{Petersen2025_DefectEng_4} developed Dis-GEN, the first generative model capable of producing symmetry-consistent disordered crystal structures with partial occupancies and vacancies, trained directly on ICSD data~\cite{Petersen2025_DefectEng_4}.
Dis-GEN is a VAE-based framework that explicitly represents occupational disorder and vacancies at the level of crystallographic sites. Disorder is encoded at each Wyckoff position together with a binary indicator specifying whether a site is ordered or disordered. This enables the generation of configurationally disordered structures that more faithfully reflect experimental crystallography.
Together, these developments mark an important shift from modeling perfect crystals toward embracing the inherent complexity of real materials by incorporating defects and disorder.
However, the widespread adoption of defect- and disorder-aware generative models remains constrained by the scarcity of data on defects and disordered materials, making this a major bottleneck for inverse design workflows.

\subsubsection*{Generative models for advanced characterization}
As chemistry and materials research laboratories become increasingly automated and autonomous, some of the traditional bottlenecks of synthesis and characterization are gradually shifting toward data analysis and interpretation.
Emerging research leverages generative models for crystal structure prediction to assist in advanced characterization and data analysis, including image and structure reconstruction during atomic-scale microscopy and ptychography~\cite{Ma2020_Micro_1, Huang2025_Micro_2, McCray2025_Micro_3, McCray2025_Micro_3}, and structure solution approaches based on powder X-ray diffraction data~\cite{guoInitioStructureSolutions2025, rieselCrystalStructureDetermination2024, Johansen2025_PXRD_1, Choudhary2025_PXRD_2}. Embedding generative models for crystal structures as components in data analysis and characterization, as well as other workflows beyond direct materials design, will become increasingly relevant with maturing and usable generative models.

\subsubsection*{Incorporating synthetic feasibility into generative models}
Conditioning generative models on synthesizability represents a key upcoming challenge, as synthesis considerations are currently addressed only in post-processing workflows (see Section~\ref{sec:post_generation}). Future progress might draw inspiration from advances in molecular generation, including the use of synthetic accessibility scores. Recent works propose analogous metrics for crystal structures, offering new opportunities to guide or condition generative models directly on synthesizability~\cite{prein2025synthesizabilityguidedpipelinematerialsdiscovery, song2025accurate}.  Moreover, emerging models that predict viable precursor combinations provide an additional pathway for integrating synthesis considerations, playing a role analogous to retrosynthesis models in molecular design~\cite{noh2025retrievalretroretrievalbasedinorganicretrosynthesis, prein2025retrorankinrankingbasedapproachinorganic}. These developments are further supported by progress in assembling synthesis-focused datasets for inorganic materials~\cite{kononova2019text}, which are essential for training and validating such synthesis-aware generative frameworks.

\subsubsection*{Model interpretability and explainability}
In recent years, the question of model interpretability has emerged as a central topic in machine learning research~\cite{longoExplainableArtificialIntelligence2024a}. The ever-increasing complexity of modern models and their resulting incomprehensible black-box character have raised concerns about user trust, technical pitfalls such as artifacts and hallucinations, and societal concerns about model bias and fairness. Research in explainable AI aims to develop methods to mitigate these risks by increasing the transparency of these black box models. While not as well explored as for predictive models, some research exists on the interpretability of generative models~\cite{schneiderExplainableGenerativeAI2024}, centered around meaningful latent space decompositions~\cite{moranInterpretableDeepGenerative2025}, identification of particularly influential training samples~\cite{nguyenRegionLevelDataAttribution2025}, and model parameters~\cite{nguyenUnveilingConceptAttribution2024}.
While these topics are being actively explored in the greater context of generative modeling, the aspect of interpretability is still notably underrepresented in crystal structure generation, aside from some isolated discussion of latent space decomposition~\cite{wangInverseDesignCatalytic2025,court3DInorganicCrystal2020}, demonstrating an important direction for future developments of the field.

\section{Summary and outlook}

Advancements in generative models for crystal structures benefited from rapid innovations in generative models in general, from VAEs and GANs to LLMs and diffusion/flow matching/BFNs. Currently, iterative generative models, such as BFNs and specifically trained transformers with crystal-structure-specific tokenizations, are on par in performance, and hybrid models are not substantially outperforming them. However, there is no reason to expect that these innovation cycles will stop, so we can expect further improvements for crystal structure generation as well, even though it is difficult to predict what the next innovations will be.
Developing faster diffusion and flow matching models that require fewer integration steps is an active area of research and will be transferable to crystal generation.
Similarly, many generative models currently rely on equivariant (graph) neural networks, which means that improvements in such architectures will also translate to improvements in crystal structure generation.
From 2016 to approximately 2022, innovations in generative models were usually adopted for molecules first before being extended to crystal structures. Since then, development cycles have shortened, and the open challenges in molecular and materials modeling have broadened. New ways of integrating synthesis, \textit{e.g.}, using GFlowNets, are under development, but inherent differences between molecules and crystalline materials are leading to parallel developments to apply such models for synthesis conditioning or even parallel structure-synthesis generation.\\

Progress in crystal representation has led to notable performance gains, primarily by embedding explicit symmetry information that models previously had to infer. Starting with the transition from Cartesian to fractional coordinates, subsequent innovations such as asymmetric unit representations and symmetry-aware tokenization for transformers have further enhanced generative fidelity.
Alternative new crystal structure representations might therefore advance the generation process without innovations on the model side. \\

Beyond models and representations, available data is a key ingredient for further progress. Most early generative models were trained on simpler subsets of the Materials Project, \textit{e.g.}, perov or MP-20. Recently, benchmarks also include more challenging datasets with larger unit cells, \textit{e.g.}, MPTS-52. 
As only ~10\% of the structures in the Materials Project have more than 52 atoms, high metrics on benchmarks like MPTS-52 will cover most of the application-relevant materials regarding unit cell size, but still exclude specific materials classes with larger unit cells, such as metal-organic frameworks. 
Scaling to larger unit cell sizes benefits from models that work on the asymmetric unit representation. It is currently unclear whether transformer-based architectures with specific tokenization related to asymmetric unit cells will have an advantage over GNN-based models, which typically require full connectivity among all atoms in the asymmetric unit cell.\\

From the application perspective, generative models should always generate materials with specific property requirements.
Most studies on generative models for crystal structures demonstrate the possibility of conditioning generation on simple scalar properties such as formation energy or band gap, or even spectral features. However, enabling targeted multi-objective generation remains an open challenge, as no general off-the-shelf frameworks or sufficiently large datasets currently exist to support it, even though classifier-free guidance emerges as an approach that works on multiple target properties.
Most papers so far report proof-of-concept studies regarding conditioning approaches, and there are only a few application studies that show that conditioning can be used to generate application-relevant materials.
Recent advances in conditional generative modeling for images and molecules~\cite{wang_training_2025}, including training-free inference-time conditioning and non-differentiable conditions, aim to eliminate the need for specialized model retraining, improving conditioning flexibility and applicability. However, this comes at the cost of shifting computational costs from training to inference time. Future adoptions of such advances into crystal generation may have the potential to make complex (multi-)property conditioning more broadly applicable in practice.
Advances in generative models conditioned on complex properties will allow their integration in characterization and data analysis workflows for autonomous labs and potentially other applications beyond materials design.\\

Moving from a complex post-processing workflow to a more end-to-end solution remains a challenge in the area of generative models for materials.
Currently, stability assessment of generated crystal structures often requires DFT-based relaxation and energy evaluation to compare with the convex hull. Advancements in the accuracy of generative models and in energy prediction models will continually decrease the need for postprocessing steps.
Despite all advances in generative models and their accuracy in benchmark tasks, as well as conditioning on application-relevant properties, there are still fundamental limitations that create a gap to experimental realization: this includes the fact that most generative models only take into account perfect crystal structures, while most application-relevant materials and their properties are at least influenced by, or even fully determined by doping, occupational and structural disorder, defects, and microstructure, posing a range of open challenges to current generative models. Furthermore, even for ideal crystal structures, integrating synthetic accessibility/synthesis prediction in generative models is an emerging topic with the potential to close the gap between progress in ML methods and transfer to experimental validation in relevant applications.

\medskip
\section*{Acknowledgement}
H.M. acknowledges financial support from the Helmholtz Project 'SolarTAP'.
L.R. acknowledges funding by the German Research Foundation (DFG) through the collaborative research center CRC 1249 'N-Heteropolycycles as Functional Materials' (SFB 1249, Project C13).
L.N.W. acknowledges funding support from the BIDMaP Postdoctoral Fellowship.
L.T. acknowledges support by the Federal Ministry of Education and Research (BMBF) under Grant No. 01DM21002A (FLAIM).
J.T. acknowledges funding from the pilot program Core-Informatics of the Helmholtz Association (HGF).
H.S. acknowledges financial support by the German Research Foundation (DFG) through the Research Training Group 2450 'Tailored Scale-Bridging Approaches to Computational Nanoscience'.
J.Ö. acknowledges funding by the Helmholtz Foundation Model Initiative supported by the Helmholtz Association.
Y.Z. acknowledges funding by the European Union (ERC, BiCMat, 101054577).
M.N. acknowledges funding from the Klaus Tschira Stiftung gGmbH (SIMPLAIX project 1).
Y.K. acknowledges funding by the German Research Foundation (DFG) within Priority Programme SPP 2331.
G.C. was funded by the U.S. Department of Energy, Office of Science, Office of Basic Energy Sciences, Materials Sciences and Engineering Division under Contract No. DE-AC0205CH11231 (Materials Project program KC23MP). 
P.F. acknowledges funding by the German Research Foundation (DFG) under Germany's Excellence Strategy via the Excellence Cluster '3D Matter Made to Order' (3DMM2O, EXC-2082/1–390761711) and by the Federal Ministry of Education and Research (BMBF) under Grant No. 01DM21001B (German-Canadian Materials Acceleration Center).

\clearpage

\printbibliography

\end{document}